\documentclass[twocolumn,showpacs,pra]{revtex4}
\usepackage{graphicx}
\usepackage{color}
\usepackage{amsmath}

\newcommand{\pd}[2]{\frac{\partial #1}{\partial #2}}
\newcommand{\pdd}[2]{\frac{\mathrm{d} #1}{\mathrm{d} #2}}

\newcommand{\pdn}[3]{\frac{\partial^{#3} #1}{\partial #2^{#3}}}

\newcommand{\operp}{\omega_{\perp}}
\newcommand{\Phidd}{\Phi_{\rm dd}}
\newcommand{\edd}{\varepsilon_{\rm dd}}
\newcommand{\Cdd}{C_{\rm dd}}
\newcommand{\vectwo}[2]{\left[ \begin{array}{c} #1 \vspace{0.2cm}\\ #2 
\end{array}\right]}
\newcommand{\matfour}[4]{\left[ \begin{array}{cc} #1 & #2 \vspace{0.2cm}\\ #3 & #4  
\end{array}\right]}
\newcommand{\br}[1]{\left(#1\right)}

\newcommand{\rvec}{\mathbf{r}}

\newcommand{\rpvec}{\mathbf{r'}}
\newcommand{\drpvec}{\mathrm{d}^3 r'}
\newcommand{\omx}{\omega_x}
\newcommand{\omy}{\omega_y}
\newcommand{\omz}{\omega_z}
\newcommand{\dn}{\delta n}
\newcommand{\phase}{S}
\newcommand{\dphase}{\delta \phase}
\newcommand{\pertop}{\mathcal{L}}
\newcommand{\tipsi}{\tilde{\psi}({\bf r})}
\newcommand{\ltsimeq}{\raisebox{-0.6ex}{$\,\stackrel
        {\raisebox{-.2ex}{$\textstyle <$}}{\sim}\,$}}
\newcommand{\gtsimeq}{\raisebox{-0.6ex}{$\,\stackrel
        {\raisebox{-.2ex}{$\textstyle >$}}{\sim}\,$}}

\newcommand\romand{\mathrm{d}}

\begin{document}

\title{Collective excitation frequencies and stationary states of trapped dipolar Bose-Einstein condensates in the Thomas-Fermi regime}

\author{R. M. W. van Bijnen$^{1,2}$, N. G. Parker$^{2, 3}$, S. J. J. M. F. Kokkelmans$^{1}$, A. M. Martin$^4$ and D. H. J. O'Dell$^2$}

\address{$^{1}$ Eindhoven University of
Technology, PO Box 513, 5600 MB Eindhoven, The Netherlands. \\
$^{2}$ Department of Physics and Astronomy, McMaster University, Hamilton, Ontario, L8S 4M1, Canada. \\
$^{3}$ School of Food Science and Nutrition, University of Leeds, LS2 9JT, United Kingdom. \\
$^{4}$ School of Physics, University of Melbourne, Parkville, Victoria 3010, Australia.}

\date{\today}

\begin{abstract}
We present a general method for obtaining the exact static solutions and collective excitation frequencies  of a trapped Bose-Einstein condensate (BEC) with dipolar atomic interactions in the Thomas-Fermi regime. The method incorporates analytic expressions for the dipolar potential of an arbitrary polynomial density profile, thereby reducing the problem of handling non-local dipolar interactions to the solution of algebraic equations. 
 We comprehensively map out the static solutions and excitation modes, including non-cylindrically symmetric traps, and also the case of negative scattering length where dipolar interactions stabilize an otherwise unstable condensate. The dynamical stability of the excitation modes gives insight into the onset of collapse of a dipolar BEC. We find that global collapse is consistently mediated by an anisotropic quadrupolar collective mode, although there are two trapping regimes in which the BEC is stable against quadrupole fluctuations even as the ratio of the dipolar to $s$-wave interactions becomes infinite. Motivated by the possibility of fragmented BEC in a dipolar Bose gas due to the partially attractive interactions, we pay special attention to the scissors modes, which can provide a signature of superfluidity, and identify a long-range restoring force which is peculiar to dipolar systems. As part of the supporting material for this paper we provide the computer program used to make the calculations, including a graphical user interface.
 \end{abstract}


\maketitle

\begin{section}{Introduction\label{SecIntroduction}}
Since the realization of atomic Bose-Einstein condensates (BECs) in 1995 \cite{BEC1995}, there has been a surge of interest in quantum degenerate gases \cite{PethickSmith,Stringari}.  Despite the diluteness of these gases, interatomic interactions play an important role in determining their properties. In the majority of experiments the dominant interactions have been isotropic and asymptotically of the van der Waals type, falling off as $1/r^{6}$. At ultracold temperatures this leads to essentially pure \textit{s}-wave scattering between the atoms. An exception to this rule is provided by gases that have significant dipole-dipole interactions \cite{Griesmaier05,Beaufils08,Fattori08,Vengalattore08}.  In comparison to van der Waals type interactions, dipolar interactions are longer range and anisotropic, and this introduces rich new phenomena. For example,  a series of experiments that have revealed the anisotropic nature of dipolar interactions are those on $^{52}$Cr BECs in an external magnetic field. These have demonstrated anisotropic expansion of the condensate depending on the direction of polarization of the atomic dipoles \cite{Stuhler05,Lahaye07}, collapse and {\it d}-wave explosion \cite{Lahaye08}, and an enhanced stability against collapse in flattened geometries \cite{Koch08}. Meanwhile, an experiment with $^{39}$K atoms occupying different sites in a 1D optical lattice has demonstrated the long-range nature of dipolar interactions in BECs through dephasing of Bloch oscillations \cite{Fattori08}. Dipolar interactions have also been shown to be responsible for the formation of a spatially modulated structure of spin domains in a $^{87}$Rb spinor BEC \cite{Vengalattore08}. 

In order to incorporate atomic interactions into the Gross-Pitaevskii theory for the condensate one should use a pseudo-potential \cite{PethickSmith,Stringari}. In the presence of both dipolar and van der Waals interactions 
the pseudo-potential can be written as the sum of two terms 
$U(\rvec)=U_{s}(\rvec)+U_{\mathrm{dd}}(\rvec)$ \cite{Goral00,Santos00,Yi01,Kanjilal07}, where $\rvec$ is the relative interatomic 
separation. The long-range dipolar interaction can be treated accurately 
within the Born approximation providing one is not close to a scattering resonance \cite{Yi01,Kanjilal07}. This 
first-order approximation means that the effective interaction is replaced by 
the potential itself. This is quite different to the shorter range van der Waals 
interaction, for which the Born approximation is not valid at low temperatures, and where one rather uses the contact potential
\begin{equation}
U_s(\rvec)=g \delta(\rvec).
\end{equation}
The coupling constant $g=4 \pi \hbar^2 a_{\mathrm{s}}/m$ is given in terms of an $s$-wave scattering length $a_{\mathrm{s}}$ and the atomic mass $m$. For the dipolar interaction, we consider two atoms whose dipoles are aligned by an external field pointing 
along the direction specified by the unit vector $\hat{{\rm e}}$. The potential is then 
given by
\begin{equation}
U_{\mathrm{dd}}(\rvec)= \frac{C_{\mathrm{dd}}}{4 \pi}\, \hat{{\rm
e}}_{i} \hat{{\rm e}}_{j} \frac{\left(\delta_{i j}- 3 \hat{r}_{i}
\hat{r}_{j}\right)}{r^{3}}. \label{eqn:U}
\end{equation}
where $C_{\rm dd}$ parameterizes the strength of the dipolar interactions, $\hat{r}$ is a unit vector in the direction of {\bf r}, and summation over repeated indices is implied.  A key figure of merit is the ratio of the two coupling strengths, defined as \cite{Giovanazzi02},
\begin{equation}
 \edd=C_{\rm dd}/3g.
\end{equation}
Dipole-dipole interactions can be either magnetic or electric in origin.
To date, the dipolar interactions seen in ultracold atom experiments \cite{Griesmaier05,Vengalattore08,Beaufils08,Fattori08} have all been magnetic dipolar interactions,  for which $C_{\rm dd}=\mu_0 d^2$, where $d$ is the magnetic dipole moment and $\mu_0$ is the permeability of free space. In terms of the Bohr magneton
$\mu_{\mathrm{B}}$, the magnetic dipole moment of a $^{52}$Cr atom is $d=6 \mu_{\mathrm{B}}$ giving $\edd \approx 0.16$ \cite{Griesmaier05}. Although this is 36 times larger than the typical value of $\edd$ found in the alkalis, it is still small. Thus, unless the system is in a configuration that makes it particularly sensitive \cite{Fattori08}, and/or is specially prepared \cite{Vengalattore08}, the magnetic dipolar interactions in the atomic gases made so far tend to be masked by stronger $s$-wave interactions. In order to make dipolar interactions in BECs more visible, the Stuttgart group have succeeded in implementing magnetic Feshbach resonances \cite{Feshbach38} in $^{52}$Cr  \cite{Koch08}. These allow $g$ to be tuned from positive to negative and even to zero.  Moreover, the sign and amplitude of the effective value of $C_{\rm dd}$ can also be tuned by rapidly rotating the external polarizing field \cite{Giovanazzi02}.  Polar molecules can have huge electric dipole moments and these systems are now close to reaching degeneracy \cite{Doyle04,Sage05,Kohler06,Ospelkaus06,Ni08,Ni10}. By appropriately tuning an external electric field a large degree of control can be exerted over these systems \cite{Ni10}.  Combined with what has already been achieved in $^{52}$Cr, one can realistically explore a large parameter space of interactions.

\begin{figure}[b]
\includegraphics[width=0.9\columnwidth]{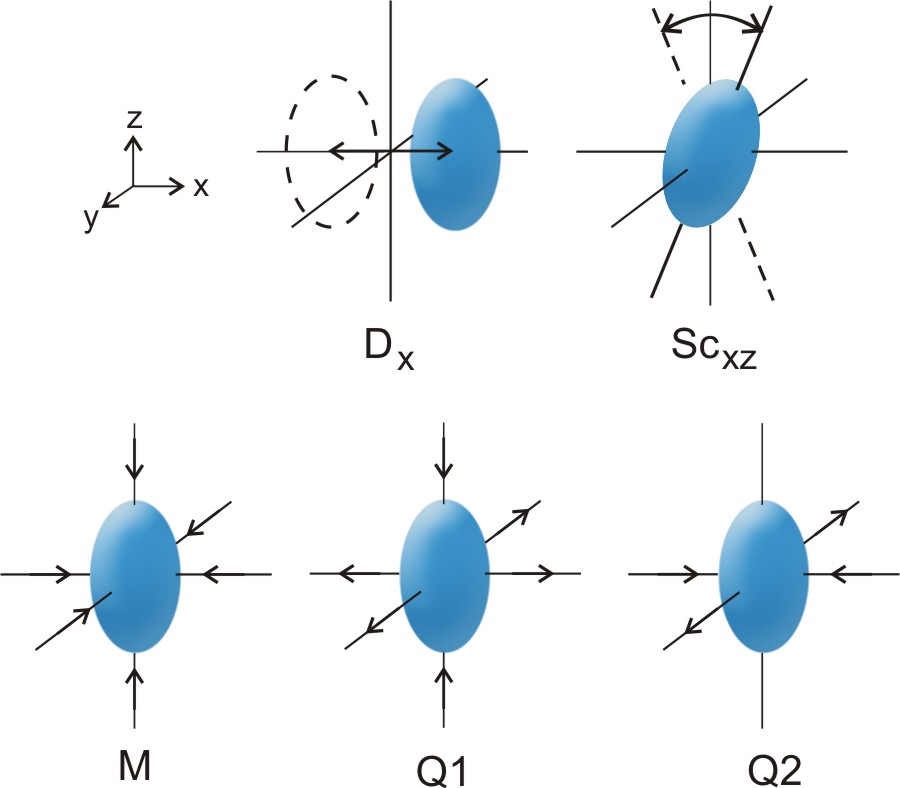}
\caption{Schematic illustration of the basic collective modes under consideration: the dipole mode $D$ (shown here in the $x$-direction $D_x$), scissors mode $Sc$ (shown here in $x-z$ plane  $Sc_{xz}$), the monopole mode $M$ and the quadrupole modes $Q_1$ and $Q_2$.  These modes are discussed in more detail in Section \ref{SecExcSpec}. \label{Modes}}
\end{figure}

The ground state of a trapped dipolar BEC  has already been investigated theoretically by a number of authors, e.g.\  \cite{Goral00,Santos00,Yi01,Kanjilal07,Goral02,ODell04,Eberlein05,Ronen06,Ronen07,Dutta07,Dutta08,Parker09}, with most studies focussing on the regime where $g \geq 0$ and $\Cdd>0$.
The presence of dipolar interactions was widely predicted to lead to certain distinctive effects, some of which have recently been seen experimentally.  For example, if the dipoles are aligned in the {\it z}-direction, then a condensate will elongate along {\it z} and become more ``cigar''-shaped, i.e.\ undergo magnetostriction, in order to benefit energetically from the attractive end-to-end interaction of dipoles. As $\edd$ is increased, for example by reducing $g$ with a Feshbach resonance, the BEC eventually becomes unstable to collapse, and this striking behavior has been realized in the experiment \cite{Lahaye08}.  Conversely, a condensate that is flattened by strong trapping along {\it z} will be mostly composed of repulsive side-by-side dipoles and so this ``pancake''-shaped geometry is more stable, as confirmed experimentally \cite{Koch08}.  In the limit that $\edd$ becomes large, but the BEC remains in the pancake configuration due to tight trapping, remarkable density wave structures have been predicted for certain regions of parameter space close to the collapse threshold \cite{Ronen07,Dutta07,Dutta08}.

In this paper we work in the Thomas-Fermi (TF) regime, which is of rather general interest because it is formally equivalent to the hydrodynamic  regime of zero-temperature superfluids \cite{Stringari96}. The TF regime may be viewed as the semiclassical approximation to the full Gross-Pitaevskii theory. A stationary condensate enters the TF regime when the zero-point kinetic energy of the atoms due to the confinement by the trap becomes negligible in comparison to the total interaction and trapping energies. For BECs with repulsive interactions in a harmonic trap this generally occurs in the large $N$ limit, where $N$ is the number of atoms. However, for dipolar BECs the picture is considerably complicated by the partially attractive and partially repulsive nature of the interactions. The question of the validity of the TF regime in dipolar BECs has been addressed in \cite{Parker08}. 

  The TF regime is theoretically simpler to handle than the full Gross-Pitaevskii theory, thereby facilitating analytical results. For example, under harmonic trapping it can be shown that the exact density profile of a dipolar condensate in the TF regime is an inverted parabola \cite{ODell04,Eberlein05}, similar to the usual \textit{s}-wave case but distorted by the magnetostriction. Furthermore, the stability of the ground state to collapse can be estimated  simply in the TF regime and reasonable agreement with experiment has been reported \cite{Koch08}.  Rotational instabilities of dipolar BECs are also amenable to analysis in the TF regime \cite{Bijnen07,Bijnen08a}. The current paper builds on these earlier works by applying the exact results available in the TF regime to collective excitations.

The excited states of a BEC can be accurately calculated within the TF regime provided they are of sufficiently long wavelength. The most basic collective excitations of  a trapped BEC are the dipole (centre-of-mass), monopole (breathing), quadrupole and scissors modes, illustrated schematically in Fig.~\ref{Modes}.  Their characterization offers important opportunities to measure interaction effects, test theoretical models, and even detect weak forces \cite{Harber05}.  Specifically, the scissors mode provides an important test for superfluidity \cite{Guery99,Marago00,Zambelli01,Cozzini03}, while the quadrupole mode plays a key role in the onset of vortex nucleation in rotating condensates   \cite{Recati01,Sinha01,Madison01,Parker06b,Bijnen07,Corro07,Bijnen08a}. An instability of the quadrupole mode is also thought to be the mechanism by which collapse of dipolar BECs proceeds when it occurs globally   \cite{Yi01,Goral02,Ticknor08,Parker09} (rather than locally \cite{Parker09}).  While the collective modes of a dipolar BEC have been studied previously \cite{Yi02,Goral02,ODell03,Santos03,ODell04,Ronen06,Giovanazzi07}, key issues remain at large, for example, the regimes of $C_{\rm dd}<0$ and $g<0$, and the behaviour of the scissors modes. This provides the motivation for the current work.

In this paper we present a general and accessible methodology for determining the static solutions and excitation frequencies of trapped dipolar BECs in the TF limit.  We explore the static solutions and the low-lying collective excitations throughout a large and experimentally relevant parameter space, including positive and negative dipolar couplings $C_{\rm dd}$, positive and negative {\it s}-wave interactions $g$, and cylindrically and non-cylindrically symmetric systems.  Moreover, our approach enables us to unambiguously identify the modes responsible for global collapse of the condensate. We would like to point out that there is a freely available MATLAB implementation of the calculations presented in this paper, complete with a graphical user interface, which can be found in the supporting material \cite{FrequenciesProgram}.

Section \ref{SecStaticSolutions} is devoted to the static solutions of the system.  Beginning with the underlying Gross-Pitaevskii theory for the condensate mean-field, we make the TF approximation and outline the methodology for deriving the TF static solutions.  We then use it to map out the static solutions with cylindrical symmetry, for both repulsive and attractive {\it s}-wave interactions, and then present an example case of the static solutions in a non-cylindrically-symmetric geometry.  
We compare to recent experimental observations where possible.  

In Section \ref{SecExcSpec} we present our methodology for deriving the excitation frequencies of a dipolar BEC.  This is an adaption of the method that Sinha and Castin  applied to standard \textit{s}-wave condensates \cite{Sinha01} where one considers perturbations around the static solutions (derived in Section \ref{SecStaticSolutions}) and employs linearized equations of motion for these perturbations.  At the heart of our approach is the exact calculation of the dipolar potential of a heterogeneous ellipsoidal BEC, performed by employing results from gravitational potential theory known in astrophysics \cite{Dyson1891,Ferrers1877,Routh, Levin71,Chandrasekhar,Rahman01} and detailed in appendices \ref{AppPotential} and \ref{AppPotential2}.

In Section \ref{SecExcCyl} we apply this method to calculate the frequencies of the important low-lying modes of the system, namely the monopole, dipole, quadrupole and scissors modes, for a cylindrically-symmetric condensate.  We show how these frequencies vary with the key parameters of the system, $\edd$ and trap ratio $\gamma$, and give physical explanations for our observations.  In Section \ref{SecExcNonCyl} we extend our analysis to non-cylindrically-symmetric BECs.  Although the parameter space of such systems is very large, we present pertinent examples.  An important feature of non-cylindrically-symmetric systems is that they support a family of scissors modes which can be employed as a test for superfluidity.  As such, in Section \ref{SecScissors}, we focus on these scissors modes and show how they vary with key parameters.  Finally, in Section \ref{SecConclusions}, we summarise our findings.  

There are three appendices included in this paper. Appendix \ref{AppCollectiveepsilondd} contains a plot of the frequencies of the collective modes of the BEC as a function of $ \edd$. Appendix \ref{AppPotential} outlines the method by which we calculate dipolar potentials due to arbitrary polynomial density distributions of atoms. This is the main technical advance of this work over our previous papers which were limited to the dipolar potentials associated with strictly paraboloidal density distributions, i.e.\ those of the same symmetry class as the static solution. In Appendix \ref{AppPotential2} we give a closed formula in terms of elliptic integrals for the dipolar potential inside a triaxial ellipsoid with a parabolic density profile. This is a special but important case of the general theory
 outlined in Appendix \ref{AppPotential}.

\end{section}

\begin{section}{Static solutions\label{SecStaticSolutions}}
\subsection{Methodology for obtaining static solutions}
At zero temperature the condensate is well-described by a mean-field order parameter, or ``wave function'', $\psi({\bf r},t)$.  This defines an atomic density distribution via $n({\bf r},t)=|\psi({\bf r},t)|^2$.  Static solutions, denoted by $\tipsi$, satisfy the time-independent Gross-Pitaevskii equation (GPE) given by \cite{Stringari}, 
\begin{eqnarray}
\left[
-\frac{\hbar^2}{2m} \nabla^2 +
V(\rvec)+\Phidd({\bf r})+g
\left|\tipsi \right|^2 \right]\tipsi=\mu \tipsi
\label{TIGPE}
\end{eqnarray}
where $\mu$ is the chemical potential of the system.  The external potential $V(\rvec)$ is typically harmonic with the general form,
\begin{equation}\label{HarmonicTrap}
V(\rvec) = \frac{1}{2} m \omega_\perp^2 \left[(1-\epsilon) x^2 + (1+\epsilon) y ^2 + \gamma^2 z^2\right].
\end{equation}
Here $\omega_\perp$ is the average trap frequency in the $x-y$ plane and the trap aspect ratio $\gamma=\omega_z/\omega_\perp$ defines the trapping in the axial ($z$) direction.  The trap ellipticity in the $x-y$ plane $\epsilon$ defines the transverse trap frequencies via $\omega_x=\sqrt{1-\epsilon} \ \omega_\perp$ and $\omega_y=\sqrt{1+\epsilon} \ \omega_\perp$. When $\epsilon=0$ the trap is cylindrically symmetric.

The $\Phidd$-term in Eq.\ (\ref{TIGPE}) is the mean-field potential arising from the dipolar interactions
\begin{equation}
 \Phidd(\rvec) =   \int n(\rpvec) U_{\mathrm{dd}}(\rvec-\rpvec) \drpvec \ .
\end{equation}
This term is a non-local functional of the density and is the source of the difficulties associated with theoretical treatments of dipolar BECs: it turns the GPE into an integro-differential equation.
A key feature of the approach taken by us in this paper is to calculate this term analytically.
To this end we express the dipolar mean-field in terms of a fictitious  electrostatic potential $\phi(\rvec)$  \cite{CraigThirunamachandran,ODell04,Eberlein05}
 \begin{equation}\label{PhiddElectrostatic}
 \Phidd(\rvec) = -\Cdd \br{\pdn{}{z}{2} \phi(\rvec) +\frac{1}{3}n(\rvec)},
 \end{equation}
where
 \begin{equation}\label{ElectrostaticPotential}
 \phi(\rvec) = \frac{1}{4\pi}\int \frac{n(\rpvec)}{|\rvec - \rpvec|} \drpvec.
 \end{equation}
$\phi(\rvec)$ satisfies Poisson's equation $\nabla^2\phi(\rvec)=-n(\rvec)$. Note that in (\ref{PhiddElectrostatic}) we have taken the dipoles to be aligned along the {\it z}-direction. The term $n(\rvec)/3$ appearing on the right hand side of (\ref{PhiddElectrostatic}) cancels the Dirac delta function which arises in the $\partial^{2} \phi(\rvec) \partial z^2$ term   \cite{CraigThirunamachandran,Hannay}. This means that $ \Phidd$ includes only the long-range ($r^{-3}$) part of the dipolar interaction, exactly as written in Equation (\ref{eqn:U}).

We assume the TF approximation where the zero-point kinetic energy of the atoms in the trap is neglected.  
Dropping the relevant $\nabla^2$-term in Eq.\ (\ref{TIGPE}) leads to
\begin{equation}\label{time_indep_GPE_harmonicV}
V(\rvec) + \Phidd(\rvec)+ g n(\rvec) = \mu.
\end{equation}
For an {\it s}-wave BEC under harmonic trapping, the exact density profile in the TF approximation is known to be an inverted parabola \cite{Stringari} with the general form,
\begin{eqnarray}
n(\rvec)=n_0\left(1-\frac{x^2}{R_x^2}-\frac{y^2}{R_y^2}-\frac{z^2}{R_z^2}\right)
\,\,\,\, {\rm for} \,\,\, n({\bf r}) \ge 0 \label{DensParabolic}
\end{eqnarray}
where $n_0=15N/(8 \pi R_x R_y R_z)$ is the central density, and $R_x, R_y,$ and $R_z$ are the condensate radii. In order to obtain the dipolar potential arising from this density distribution, one must find the corresponding electrostatic potential of  Eq.~(\ref{ElectrostaticPotential}). References \cite{ODell04,Eberlein05} follow this procedure, and arrive at the remarkable conclusion that the dipolar potential $\Phidd$ is also parabolic. Therefore, a parabolic density profile is also an exact solution of the time-independent TF Equation (\ref{time_indep_GPE_harmonicV})  even in the presence of dipolar interactions.  In Section \ref{SecExcSpec} and Appendices \ref{AppPotential} and \ref{AppPotential2} we point out that this result can be extended using results from 19th century gravitational potential theory \cite{Dyson1891, Ferrers1877, Levin71} to arbitrary polynomial densities yielding polynomial dipolar potentials of the same degree. For the parabolic density profile at hand, the internal dipolar potential is given by \cite{Eberlein05, Bijnen07}
\begin{eqnarray}
&&\Phidd(\rvec) = - g \edd n(\rvec) +\frac{3 g \edd n_0 \kappa_x \kappa_y}{2} \nonumber\\ 
&&\times \left[\beta_{001}-\br{\beta_{101}x^2 + \beta_{011}y^2+3\beta_{002}z^2} R_z^{-2} \right] \label{PhiddParabolic}
\end{eqnarray}
where $\kappa_x = R_x / R_z$ and $\kappa_y = R_y / R_z$ are the aspect ratios of the condensate, and
\begin{equation}\label{Betalmn}
\beta_{ijk} = \int_0^{\infty} \frac{\mathrm{d}s}{(\kappa_x^2+s)^{i+\frac{1}{2}} (\kappa_y^2+s)^{j+\frac{1}{2}} (1+s)^{k+\frac{1}{2}}},
\end{equation}
where $i,j,k$ are integers. Explicit expressions for $\beta_{001},\beta_{101},\beta_{011}$, and $\beta_{002}$  in terms of elliptic integrals are given in Appendix \ref{AppPotential2}.
Note that for a cylindrically-symmetric trap $\epsilon=0$, the static condensate profile is also cylindrically-symmetric with aspect ratio $\kappa_x=\kappa_y=:\kappa$.   In the cylindrically-symmetric case the integrals $\beta_{ijk}$ of Eq.~(\ref{Betalmn}) can be evaluated in terms of the $_2F_1$  Gauss hypergeometric function \cite{GradshteynRyzhik,AbramowitzStegun} for \emph{any} $i,j,k$
\begin{equation}\label{BetalmnCyl}
\beta_{ijk} = 2 \frac{\ _2F_1\br{k+\frac{1}{2}, 1; i+j+k+\frac{3}{2};1-\kappa^2}}{(1 + 2i + 2j + 2k)\kappa^{2(i + j)}}.
\end{equation}

For the parabolic density profile of Eq.\ (\ref{DensParabolic}), the  TF Eq.\ (\ref{time_indep_GPE_harmonicV}) becomes
\begin{eqnarray}
\mu &=& 3g \edd \frac{n_0 \kappa_x \kappa_y}{2R_z^2}\left[R_z^2\beta_{001} - \beta_{101} x^2 -\beta_{011}y^2 - 3\beta_{002}z^2\right] \nonumber \\
&+&V(\rvec) + (1-\edd)\frac{g n_0}{R_z^2}\br{R_z^2 - \frac{x^2}{\kappa_x^2} - \frac{y^2}{\kappa_y^2} - z^2} \label{time_indep_GPE}.
\end{eqnarray}
Inspection of the coefficients of $x^2, y^2$ and $z^2$ leads to three self-consistency relations, given by
\begin{eqnarray}
\kappa_x^2 &=& \frac{\omz^2}{\omx^2}\frac{1+\edd\left(\frac{3}{2}\kappa_x^3\kappa_y \beta_{101}-1\right)}{1-\edd\left(1-\frac{9 \kappa_x \kappa_y}{2}\beta_{002}\right)} \label{kappax}, \\
\kappa_y^2 &=& \frac{\omz^2}{\omy^2}\frac{1+\edd\left(\frac{3}{2}\kappa_y^3\kappa_x \beta_{011}-1\right)}{1-\edd\left(1-\frac{9 \kappa_x \kappa_y}{2}\beta_{002}\right)} \label{kappay}, \\
R_z^2&=&\frac{2gn_0}{m\omz^2}\left[1-\edd\left(1-\frac{9 \kappa_x \kappa_y}{2}\beta_{002}\right)\right] \label{Rz}.
\end{eqnarray}
Solving Eqs.~(\ref{kappax})-(\ref{Rz}) gives the exact static solutions of the system in the TF regime.   

The energetic stability of the condensate is determined by the TF energy functional
\begin{equation}
E=\int \left(V(\rvec)+ \frac{1}{2}\Phi_{\mathrm{dd}}(\rvec)+\frac{1}{2}g n(\rvec)\right)n(\rvec) \mathrm{d}^{3}r.
\end{equation}
Inserting the parabolic density profile (\ref{DensParabolic}) yields an energy landscape
\begin{eqnarray}
E &=& \frac{15 N^2 g}{28 \pi \kappa_x \kappa_y R_z^3}\left[ (1-\edd)\right. \nonumber \\
 &+& \left. \frac{3}{8}\kappa_x \kappa_y \edd \br{7 \beta_{001} - 3 \beta_{002} - \kappa_x^2\beta_{101} - \kappa_y^2 \beta_{011}} \right] \nonumber \\
 &+& \frac{N}{14}m R_z^2 \br{\kappa_x^2 \omega_x^2 + \kappa_y^2 \omega_y^2 +\gamma^2}. \label{eq:TFEnergy}
\end{eqnarray}
Static solutions correspond to stationary points in the energy landscape.  If the stationary point is a local minimum in the energy landscape, it corresponds to a physically stable solution.  However, if the stationary point is a maximum or a saddle point, the corresponding solution will be energetically unstable.  The nature of the stationary point can be determined by performing a second derivative test on Eq.~(\ref{eq:TFEnergy}) with respect to the variables $\kappa_x, \kappa_y,$ and $R_z$.  This leads to 6 lengthy equations that  will not be presented here. Note that this only determines whether the stationary point is a local minimum \textit{within the class of parabolic density profiles}. 
In other words, with the three variables $\kappa_x, \kappa_y,$ and $R_z$ we are only able to determine stability against ``scaling'' fluctuations, so named because they  correspond to a rescaling of the static solution \cite{Kagan96,Castin96}. However,
the class of scaling fluctuations includes important low-lying shape oscillations such as the monopole and quadrupole modes.
%REPHRASE THE FOLLOWING: 
Although higher order (beyond quadrupole) modes can become unstable in certain regimes, as a criterion of stability we will use the local minima of (\ref{eq:TFEnergy}). This assumption is supported by the recent experiments by Koch {\it et al.} \cite{Koch08}, where dipolar BECs were produced with $\edd > 1$ that were stable over significant time-scales.

%Typically, higher order modes require symmetry breaking and might take a while to be excited. Also, in light of the recent experiments by Koch et al \cite{Koch08}, we deem the $\edd > 1$ and $\edd < -0.5$ ($g>0$) parameter range metastable. BLA BLA.

\subsection{Cylindrically-symmetric static solutions for $g>0$, and the critical trap ratios $\gamma_{\rm crit}^+$ and $\gamma_{\rm crit}^-$}\label{SubSecCyl1}
We have obtained the static solutions for a cylindrically-symmetric BEC by solving Eqs.~(\ref{kappax}) to (\ref{Rz}) numerically.  The solutions behave differently depending on whether the {\it s}-wave interactions are repulsive or attractive.   We begin by considering the $g>0$ case.  The ensuing static solutions, characterised by their aspect ratio $\kappa$, are presented in Fig.~\ref{FigStaticCyl1} as a function of $\edd$ with each line representing a different trap ratio $\gamma$.   While the TF solutions in the regime $\edd>0$ have been discussed previously \cite{ODell04,Eberlein05},  the regime of $\edd<0$ has not been studied.  Be aware that when we fix $g>0$, the regime $\edd<0$ (left hand side of Fig.\ \ref{FigStaticCyl1}) corresponds to $C_{\rm dd}<0$ where the dipolar interaction is reversed, repelling along $z$ and attracting in the transverse direction.  This can be achieved by rapid rotation of the field aligning the dipoles about the {\it z}-axis  \cite{Giovanazzi02}.  

Before we examine the question of stability, let us first interpret the structure of the solutions shown in Fig.\ \ref{FigStaticCyl1}. Imagine an experiment in which the magnitude of $\edd$ is slowly increased from zero. At $\edd=0$ we have purely $s$-wave interactions and all solutions have the same aspect ratio as the trap, i.e. $\kappa=\gamma$. As $\edd$ is increased above zero $\kappa$ decreases so that $\kappa < \gamma$ for all solutions. This is because standard magnetostriction causes dipolar BECs to be more cigar-shaped than their $s$-wave counterparts. Conversely, if $\edd$ is made negative then $\kappa$ increases so that $\kappa> \gamma$ for all solutions. This is because when $C_{\mathrm{dd}}<0$  we have non-standard (reversed) magnetostriction which leads to a more pancake shaped BEC.

Consider now the stability of the solutions, beginning with the range $-1/2 <\edd<1$ [white region in Fig.~(\ref{FigStaticCyl1})].
We find that the energy landscape (\ref{eq:TFEnergy}) has only one stationary point, namely a global energy minimum, and it occurs at finite values of the radii $R_{x} (= R_{y})$, and $R_{z}$. This global minimum persists for all trap ratios
(outside of the range $-1/2 <\edd<1$ the existence of stable static solutions depends on $\gamma$).
Thus, in the range $-1/2 <\edd<1$ the static TF solution is stable against  scaling fluctuations. Other classes of perturbation could lead to instability, but  there is good reason to believe that in this range the parabolic solution is stable against these too. Take, for example, phonons, i.e. local density perturbations. These have a character that can be considered opposite to the global motion involved in scaling oscillations. The local character of phonons means that considerable insight can be gained from the limiting case of a homogeneous dipolar condensate.  The energy of a plane wave perturbation (phonon) with momentum $p$ is given by the Bogoliubov energy $E_{\mathrm{B}}$ \cite{Goral00},
\begin{equation}
E_{\mathrm{B}}^2=\left(\frac{p^{2}}{2m}\right)^{2}
+2gn\left\{1+\edd \left(3 \cos^{2} \theta -1 \right) \right\}\frac{p^{2}}{2m},
\label{eq:bogdispersion}
\end{equation}
where $\theta$ is the angle between the momentum of the phonon and the polarization direction.  
The perturbation evolves as $\sim \exp(i E_{\rm B}t/\hbar)$ and so when $E_{\rm B}^2<0$ the perturbations grow exponentially, signifying a dynamical instability.  Dynamical stability requires that $E_{\rm B}^2>0$ which, for $g>0$, corresponds to the requirement that $[1+\edd(3\cos^2 \theta-1)] \geq 0$ in Eq.~(\ref{eq:bogdispersion}).  This leads once again to precisely the stability condition $-1/2<\edd<1$.  
%For trapped condensates, variations of $\edd$ induce a modification of the condensate aspect ratio $\kappa$.  Note that for $\edd=0$, $\kappa=\gamma$.  For $\edd>0$, the dipolar interaction is attractive along $z$ and repulsive perpendicular to it.  Hence, as $\edd$ is increased the condensate elongates along $z$, i.e., $\kappa$ decreases, in order to minimize the interaction energy.  Conversely, for $\edd<0$, the effective dipolar interaction is repulsive along $z$ and attractive transversely.  Hence as $|\edd|$ increases the system flattens and $\kappa$ increases.  These effects can be likened to magnetostriction. 

\begin{figure}[t]
\includegraphics[width=0.9\columnwidth]{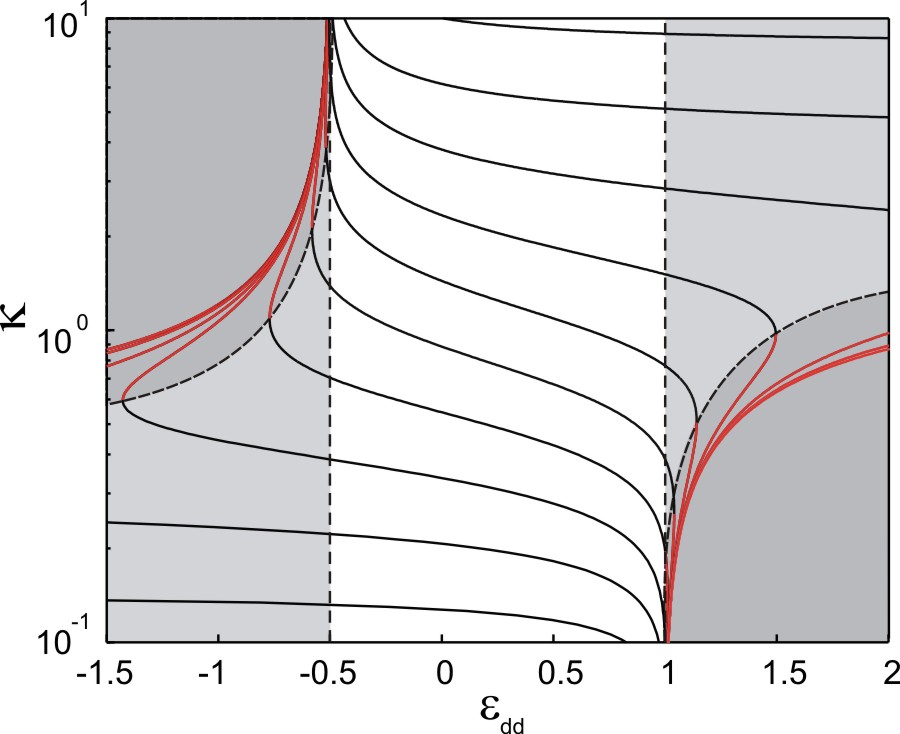}
\caption{(Color online) Aspect ratio $\kappa$ of the $g>0$ cylindrically-symmmetric static solutions as a function of $\edd$ according to Eqs.~(\ref{kappax})-(\ref{Rz}).  Note that $\edd<0$ corresponds to $\Cdd<0$.  The solid lines indicate the static solutions for specific trap ratios $\gamma$ which are equally spaced on a logarithmic scale in the range $\gamma=[0.1,10]$, with black/red lines correspond to minimum/saddle points in the energy landscape.  The parameter space of global, metastable and unstable solutions is denoted by white, light grey and dark grey regions, respectively.}  \label{FigStaticCyl1}
\end{figure}

Outside of the regime $-1/2<\edd<1$ the {\em global} energy minimum of the TF system is a collapsed state where at least one of the radii is zero, just like in the uniform dipolar BEC case.  However, unlike the uniform case, in the presence of a trap the energy functional can also support a {\em local} energy minimum corresponding to a metastable solution [light grey region in Fig.~\ref{FigStaticCyl1}].  The existence of a metastable solution means there must also be a saddle point connecting the metastable solution to the collapsed state and this is indicated by the dark grey region in Fig.~(\ref{FigStaticCyl1}).   

In general, the occurence of metastable solutions depends sensitively on $\edd$ and $\gamma$.  Remarkably, however, there are two critical trap ratios, $\gamma_{\rm crit}^+=5.17$ and $\gamma_{\rm crit}^-=0.19$,  beyond which the BEC is stable against scaling fluctuations even as the strength of the dipolar interactions becomes infinite. 
First consider $\edd>1$, for which there is a susceptibility for collapse towards an infinitely narrow line of end-to-end dipoles ($R_{x} = R_{y} \rightarrow 0$). Providing $\gamma > \gamma_{\rm crit}^+$, i.e.\ if the trap is pancake enough, condensate solutions metastable against scaling fluctuations persist even as  $\edd\rightarrow \infty$ \cite{Santos00,Yi01,Eberlein05}. Referring to Fig.\ \ref{FigStaticCyl1}, these curves are located in the upper right hand portion of the plot and asymptote to horizontal lines as $\edd$ is increased (see Fig.\  3 in \cite{Eberlein05} for a plot which extends $\edd$ to much higher values than shown here so that this behavior is clearer).   However, if the trap is not pancake-shaped enough, i.e.\ $\gamma < \gamma_{\rm crit}^+$, then as $\edd$ is increased from zero the local energy minimum eventually disappears and no stable solutions exist.  Referring again to Fig.\ \ref{FigStaticCyl1}, these are the curves that turn over as $\edd$ is increased, and in so doing enter the dark grey region.  Second, consider $\edd<-0.5$, for which the system is susceptible to collapse into an infinitely thin pancake of side-by-side dipoles ($R_z \rightarrow 0$).  If the trap is sufficiently cigar-like with $\gamma<\gamma_{\rm crit}^-$ collapse via scaling oscillations is suppressed even in the limit $\edd\rightarrow -\infty$. These curves are located in the lower left hand portion of Fig.\ \ref{FigStaticCyl1} and asymptote to horizontal lines.   However, if the trap is not cigar-shaped enough, i.e.\ $\gamma>\gamma_{\rm crit}^-$, then for sufficiently large and negative $\edd$ the metastable solution disappears, bending upwards to enter the dark grey region on the left hand portion of Fig.\ \ref{FigStaticCyl1} and the system becomes unstable to collapse.

In a recent experiment Lahaye {\it et al.}\ \cite{Lahaye08} measured the aspect ratio of the dipolar condensate over the range $0 \ltsimeq \edd \ltsimeq 1$, using a Feshbach resonance to tune $g$, and found very good agreement with the TF predictions.  Similarly,  Koch {\it et al.}\ \cite{Koch08} observed the threshold for collapse in a $\gamma=1$ system to be $\edd\approx 1.1$, in excellent agreement with the TF prediction of $\edd = 1.06$.  Using various trap ratios, it was also found that collapse became suppressed in flattened geometries and the critical trap ratio was observed to exist in the range $\gamma_{\rm crit}^+\approx 5-10$, which is in qualitative agreement with the TF predictions. 

\subsection{Cylindrically-symmetric static solutions for $g<0$, and the nature of dipolar stabilization}\label{SubSecCyl2}

\begin{figure}[t]
\includegraphics[width=0.9\columnwidth]{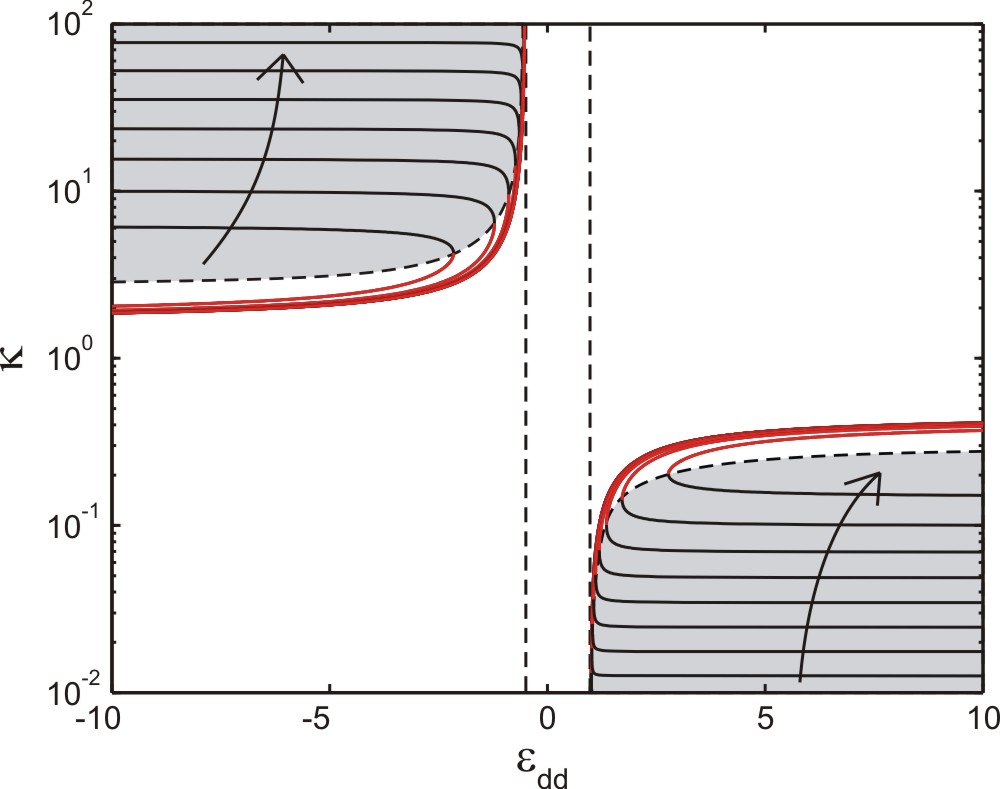}
\caption{(Color online) Aspect ratio $\kappa$ of the $g<0$ cylindrically-symmetric static solutions as a function of $\edd$.  Note that the regime of $\edd>0$ corresponds to $\Cdd<0$.  The lines denote static solutions for specific trap ratios $\gamma$, equally spaced on a logarithmic scale in the ranges $\gamma=[0.010,\gamma_{\rm crit}^-]$ (lower right set of curves) and $\gamma=[\gamma_{\rm crit}^+, 100]$ (upper left set of curves).  Arrows indicate direction of increasing $\gamma$. The light grey region and black lines correspond to minimum points, while red lines correspond to saddle points in the energy landscape. At the extreme left and right hand sides of the figure the stable solutions become horizontal lines as they tend asymptotically to the trap aspect ratio $\kappa \rightarrow \gamma$ (see text).}  \label{FigStaticCyl2}
\end{figure}

We now consider the case of attractive {\it s}-wave interactions $g<0$. Negative values of $g$ can be achieved using a Feshbach resonance, as implemented in a $^{52}$Cr BEC in \cite{Koch08}.  The static solutions are presented in Fig.~\ref{FigStaticCyl2}.    Be aware that because $g<0$, $\edd<0$ ($\edd>0$) now corresponds to $C_{\rm dd}>0$ ($C_{\rm dd}<0$).    The stability diagram differs greatly from the $g>0$ case and, in particular, no TF solutions exist  in the range $-1/2<\edd<1$.  Nevertheless, TF  solutions can exist outside of this range in regions of parameter space determined by the two critical trap ratios $\gamma_{\rm crit}^-$ and $\gamma_{\rm crit}^+$ introduced in the previous section. We find that for $\edd>0$ solutions only exist for significantly cigar-shaped geometries with $\gamma<\gamma_{\rm crit}^- = 0.19$, while for $\edd<0$ solutions only exist for significantly pancake-shaped geometries with $\gamma>\gamma_{\rm crit}^+=5.17$.  Furthermore, the attractive {\it s}-wave interactions always cause the global minimum to be a collapsed state.  This means that static solutions are only ever metastable (light grey region in Fig.~\ref{FigStaticCyl2}).  

Again, valuable insight can be gained by considering the Bogoliubov spectrum (\ref{eq:bogdispersion}), this time with $g<0$.  Firstly, for the purely {\it s}-wave case we recall the well-known result \cite{Stringari} that a homogeneous attractive BEC is always unstable to collapse.  With dipolar interactions the uniform system is stable to axial perturbations ($\theta=0$) for $\edd<-1/2$ and to radial perturbations ($\theta=\pi/2$) for $\edd>1$.  This is the exact opposite of the $g>0$ case and corroborates the lack of solutions given by the TF equations for $-1/2<\edd<1$.  Of course, $\edd<-1/2$ and $\edd>1$ cannot be simultaneously satisfied and so a uniform dipolar system with $g<0$ is always unstable.   However, when the system is trapped the condensate can be stabilized even in the TF regime. The mean dipolar interaction depends on the condensate shape and can become net repulsive in cigar-shaped systems when $\edd>0$ (for which $C_{\rm dd}<0$), and in pancake-shaped systems when $\edd<0$ (for which $C_{\rm dd}>0$).  Remarkably, in these cases it is the dipolar interactions that stabilize the BEC against the attractive {\it s}-wave interactions and lead to the regions of metastable static solutions observed in Fig.\ \ref{FigStaticCyl2}. Without the dipolar interactions the BEC would collapse.

Although our model predicts that no solutions exist for $-1/2<\edd<1$, it is well-known that stable condensates with purely attractive {\it s}-wave interactions can exist.  Zero-point motion of the atoms (ignored in the TF model) induced by the trapping potential stabilises the condensate up to a critical number of atoms or interaction magnitude \cite{Stringari}.  One can expect, therefore, that for a finite number of atoms the presence of zero-point motion enhances the stability of the condensate beyond the TF solutions.
Koch {\it et al.}\ \cite{Koch08} have produced a dipolar condensate with $g<0$ and reported the onset of collapse for $\edd \gtsimeq -7$ in a trap with $\gamma=10$. For this trap the TF static solutions disappear for $\edd \gtsimeq -1.5$.  The inclusion of zero-point motion cannot explain this discrepancy between theory and experiment since 
it should increase the critical value of $\edd$ beyond $-1.5$, not decrease it. Furthermore, including the zero-point motion by using a gaussian ansatz leads to an almost identical prediction \cite{Koch08}. One possible explanation of the discrepancy is that the dominant dipolar interactions may lead to significant deviations of the density profile from a single-peaked inverted parabola/gaussian profile, for example, Ronen {\it et al.} \cite{Ronen07} have predicted bi-concave density structures, albeit in the different regime of $\edd \rightarrow \infty$.  
%Experimental limitations or measurement errors may also be responsible.  

The metastable TF solutions shown in Fig.\ \ref{FigStaticCyl2} 
have a  counter-intuitive dependence upon $\edd$. Take, for example, the family of metastable solutions (black curves) in the lower right hand portion of the figure. We see that as $\edd$ increases $\kappa$ decreases (condensate becomes more cigar-shaped). This is in contradiction to what one might naively expect because on this side of the figure $C_{\rm dd}<0$, and so the dipolar interaction has an energetic preference for dipoles sitting side-by-side not end-to-end! In order to appreciate what is happening in this region of Fig.\ \ref{FigStaticCyl2}, observe that for each value of $\edd$ there is a critical value of the condensate aspect ratio $\kappa$ below which the system is metastable, and above which it is unstable.  As $\edd$ is increased from this point the net repulsive
dipolar interactions favor elongating the BEC  so that atoms sit further from each other, thereby lowering the interaction energy and decreasing $\kappa$.  In the limit $\kappa \rightarrow 0$ one can show that the dipolar mean-field potential tends to  $\Phi_{\mathrm{dd}} = - g \edd n(\rvec)$ \cite{Parker08}, i.e.\ it behaves like a spherically-symmetric contact interaction which is repulsive when $g < 0$ and $\edd > 0$. This means that when $\edd$ is increased in a strongly cigar-shaped configuration the condensate aspect ratio tends asymptotically towards that of the trap $\kappa \rightarrow \gamma$, as it must for a system with net-repulsive spherically-symmetric contact interactions. This behavior can be seen in Fig.\  \ref{FigStaticCyl2}  where the black curves all tend to straight lines as $\edd$ is increased, and the asymptotic value of $\kappa$ they tend to is exactly the trap aspect ratio $\gamma$.

A parallel argument holds for the upper left hand portion of 
Fig.\ \ref{FigStaticCyl2} where the condensate is quite strongly pancake-shaped ($\gamma >\gamma_{\rm crit}^+$): in the limit $\kappa \rightarrow \infty$ one can show that the dipolar mean-field potential tends to  $\Phi_{\mathrm{dd}} = 2 g \edd n(\rvec)$ \cite{Parker08}, i.e.\ it behaves like a spherically-symmetric contact interaction which is repulsive when $g < 0$ and $\edd < 0$. In this portion of the figure one therefore also finds that as $\vert \edd \vert  \rightarrow \infty$ the condensate aspect ratio tends asymptotically towards that of the trap $\kappa \rightarrow \gamma$.

It is tempting to conclude that the collapse that occurs as the strength of the dipolar interactions is reduced relative to the $s$-wave interactions is an ``$s$-wave collapse'' of the type encountered in
BECs with attractive purely $s$-wave interactions, which typically occur through an unstable monopole mode \cite{Sackett98}. However, from Fig.\ \ref{FigStaticCyl2} 
we see that the magnitude of the dipolar interaction is always finite at the collapse point. Furthermore, we shall find in subsequent sections that it is always a quadrupole mode that is responsible for collapse in a TF dipolar BEC. Collapse via a quadrupole mode 
has a 1D or 2D character, depending on the sign of $C_{\rm dd}$ \cite{Parker09}, and is distinct from collapse via the monopole mode which has a 3D character.

Having indicated how the static solutions behave for attractive {\it s}-wave interactions $g<0$, for the remainder of the paper we will concentrate (although not exclusively) on the more common case of repulsive {\it s}-wave interactions. 

\subsection{Non-cylindrically-symmetric static solutions}\label{SubSecAnisotropic}

\begin{figure}[h]
\includegraphics[width=0.99\columnwidth]{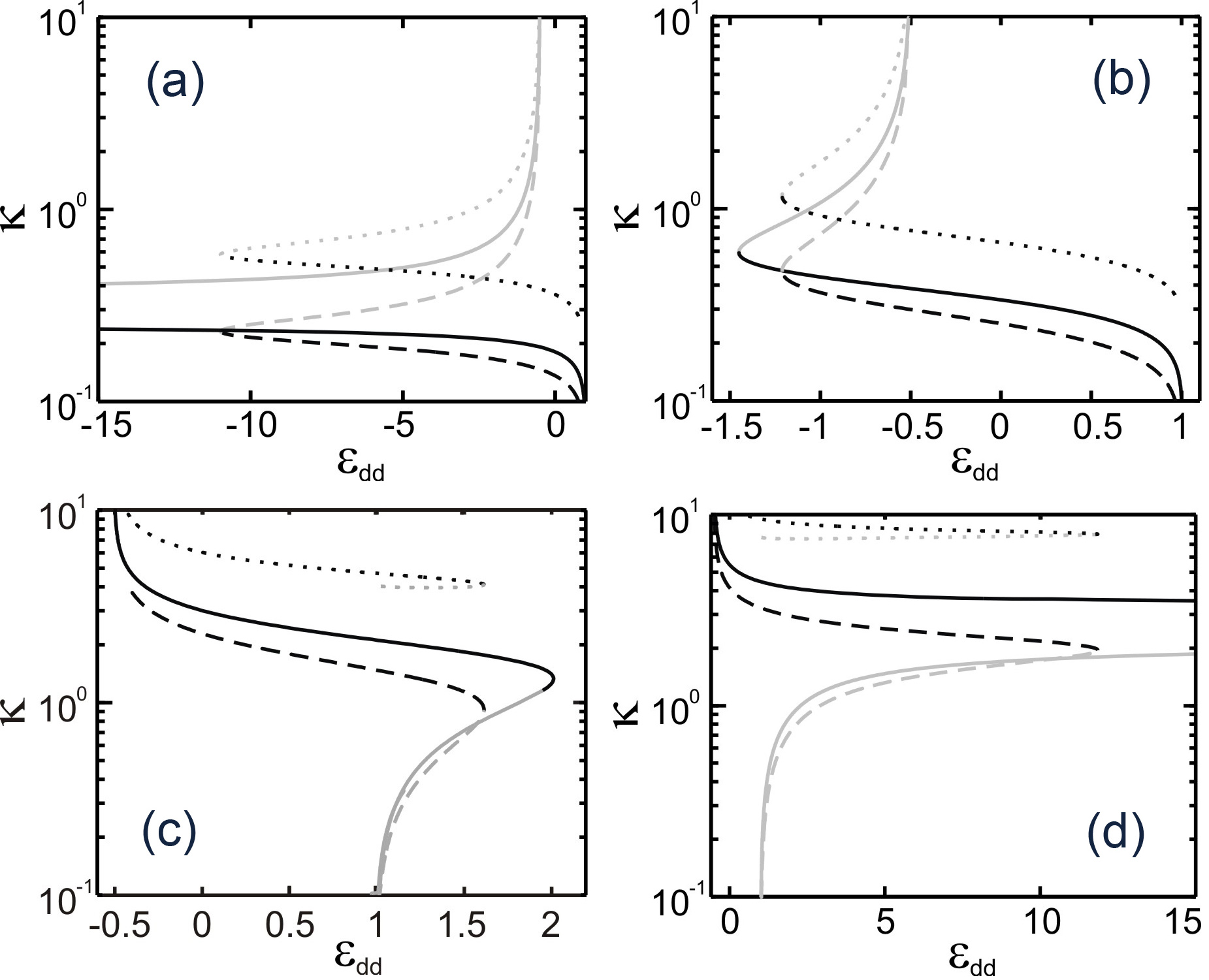}
\caption{Stable static solutions, characterised by the aspect ratios $\kappa_x$ (dotted lines) and $\kappa_y$ (dashed lines), in a non-cylindrically-symmetric trap with ellipticity $\epsilon=0.75$ and (a) $\gamma=0.18$, (b) $\gamma=0.333$, (c) $\gamma=3$ and (d) $\gamma=5.5$.  Stable (unstable) static solutions are indicated by black (grey) lines.  The corresponding static solutions for $\epsilon=0$ are indicated by solid lines.\label{FigStaticAnisotropic}}
\end{figure}

We now consider the more general case of a non-cylindrically-symmetric system for which the trap ellipticity $\epsilon$ is finite and $\kappa_x$ and $\kappa_y$ typically differ.  
Note that we perform our analysis of non-cylindrically-symmetric static solutions for repulsive {\it s}-wave interactions $g>0$.  In Fig.~\ref{FigStaticAnisotropic} we show how $\kappa_x$ and $\kappa_y$ vary as a function of $\edd$ in a non-cylindrically-symmetric trap.  Different values of trap ratio are considered and generic qualitative features exist.  The splitting of $\kappa_x$ and $\kappa_y$ is evident, with $\kappa_x$ shifting upwards and $\kappa_y$ shifting downwards in comparison to the cylindrically-symmetric solutions.  Furthermore, the branches become less stable to collapse.  For example, for $\gamma=0.18 < \gamma_{\rm crit}^{-}$ (Fig. \ref{FigStaticAnisotropic}(a)), in the cylindrically-symmetric system there exist stable solutions for $\edd \rightarrow -\infty$, but in the anisotropic case, stationary solutions only exist up to $\edd \simeq -11$. 

We already noted in the introduction that for a cylindrically symmetric dipolar BEC magnetostriction causes the radial \textit{vs} axial aspect ratio $\kappa=R_{x}/R_{z}$ to differ from the trap ratio $\gamma$, in contrast to a pure $s$-wave BEC for which $\kappa=\gamma$. It is therefore interesting to note that we find that when the trap is not cylindrically symmetric a dipolar BEC also has an ellipticity in the $xy$-plane which differs from that of the trap, although the deviation is generally small. This occurs despite the fact that dipolar interactions are radially symmetric.

\end{section}

\begin{section}{Calculation of the excitation spectrum \label{SecExcSpec}}
Now that we have exhibited some of the features of the static solutions in the TF regime, we wish to  determine their excitation spectrum. The methods which have been previously used for finding the excitation spectrum of a dipolar BEC include:
i) A variational approach applied to a gaussian approximation for the BEC density profile  \cite{Perez96,Yi01,Yi02,Goral02}. This allows one to derive equations of motion for the  widths of the gaussian.  ii)  Using the equations of dissipationless hydrodynamics, namely the continuity and Euler equations, to obtain equations of motion for the TF radii    \cite{ODell04,Giovanazzi07}. This method is exact in the TF limit (recall that the TF regime is mathematically identical to the hydrodynamics of superfluids at zero temperature).  iii) Solving the full Bogoliubov equations \cite{ODell03,Santos03,Ronen06}.  iv) Solving for the time evolution of the full time-dependent GPE under well-chosen peturbations \cite{Yi01,Yi02,Goral02}.

Methods i) and ii) are simple but yield only the three lowest energy collective modes (the monopole and two quadrupole modes). However, in the pure $s$-wave case these methods do have the advantage of giving analytic expressions for the frequencies, and in the dipolar case the frequencies are given by the solution of the algebraic equations (\ref{kappax}--\ref{Rz}), which are simple to solve. This is to be contrasted with the other methods which, although more general, require much more sophisticated numerical approaches. Furthermore, the non-local nature of the dipolar interactions make numerical calculations considerably more intensive than their $s$-wave equivalents. Therefore, the approach we adopt here is semi-analytic, incorporating analytic results for the non-local dipolar potential, thereby reducing the problem to the solution of (local) algebraic equations.

In our approach we generalize the methodology previously applied by Sinha and Castin \cite{Sinha01} to pure $s$-wave BECs, where linearized equations of motion are derived for small perturbations about the mean-field stationary solution.  One strength of this method, in contrast to some of those mentioned above, is that it is trivially extended to arbitrary modes of excitation and unstable modes/dynamical instability. For example, extension of the variational approach to higher-order modes (e.g., to consider the scissors modes of an {\it s}-wave BEC \cite{Khawaja01}) requires that this is ``built-in'' to the variational ansatz itself.  We outline our approach below.

The dynamics of the condensate wave function $\psi(\rvec,t)$  is described by the time-dependent Gross-Pitaevskii equation,
\begin{eqnarray}
i \hbar \frac{\partial \psi}{\partial t}=  \left[-\frac{\hbar^2}{2m}\nabla^2+V+\Phidd+g
\left|\psi \right|^2 \right]\psi
\label{GPE}, 
\end{eqnarray}
where, for convenience, we have dropped the arguments $\rvec$ and $t$.
By expressing $\psi$ in terms of its density $n$ and phase $\phase$ as,
\[
\psi = \sqrt{n} \mathrm{e}^{\mathrm{i} \phase},
\]
one obtains from Eq.\ (\ref{GPE}) the well-known hydrodynamic equations,
\begin{eqnarray}\label{cont_eq}
\pd{n}{t} = - \frac{\hbar}{m}\nabla \cdot \br{n \nabla \phase}\\
\label{motion_eq_phase}
\hbar\pd{\phase}{t} = - \frac{\hbar^2}{2m} |\nabla\phase|^2 - V - g n - \Phidd.
\end{eqnarray}
We have dropped the term $(\hbar^2/2m\sqrt{n})\nabla^2\sqrt{n}$ arising from density gradients - this is synonymous with making the TF approximation \cite{Stringari}.
Note that static solutions satisfy the equilibrium conditions $
\partial n/\partial t = 0$ and $\partial \phase/\partial t = -\mu/\hbar$.

We now consider small perturbations of the density and phase, $\dn$ and $\dphase$, to static solutions, and linearize the hydrodynamic equations (\ref{cont_eq}, \ref{motion_eq_phase}). The dynamics of the perturbations are then described as
\begin{equation}
\pd{}{t}\vectwo{\dphase}{\dn} = \pertop \vectwo{\dphase}{\dn} \label{perturbations},
\end{equation}
where
\begin{equation}\label{pertop}
\pertop = -\matfour{0}{g (1 + \edd K) / m }{\nabla \cdot n_0 \nabla}{0},
\end{equation}
and the operator $K$ is defined as
\begin{eqnarray}
(K \dn)(\rvec) = -3 \pdn{}{z}{2} \int \frac{\dn(\rpvec) \drpvec}{4 \pi|\rvec - \rpvec|} - \dn(\rvec) \label{Koperator}.
\end{eqnarray}
The integral in the above expression is carried out over the domain where the unperturbed density of Eq.\ (\ref{DensParabolic}) satisfies $n > 0$, that is, the general ellipsoidal domain with radii $R_x, R_y, R_z$. Extending the integration domain to the region where $n + \dn > 0$ would only add $\mathcal{O}(\dn^2)$ effects, since it is exactly in this extended domain that $n = \mathcal{O}(\dn)$, whereas the size of the extension is also proportional to $\dn$. Clearly, to first order in $\dn$, the quantity $\edd K \dn$ is the dipolar potential associated with the density distribtution $\dn$.
To obtain the global shape excitations of the BEC one has to find the eigenfunctions $\dn, \delta \phase$ and eigenvalues $\lambda$ of operator $\pertop$ of Eq.~(\ref{pertop}).  
For such eigenfunctions equation (\ref{perturbations}) trivially yields an exponential time evolution of the form $\sim\exp(\lambda t)$. When the associated eigenvalue $\lambda$ is imaginary, the eigenfunction corresponds to a time-dependent oscillation of the BEC. However, when $\lambda$ posesses a positive real part, the eigenfunction represents an unstable excitation which grows exponentially.  Such dynamical instabilities are an important consideration, for example in rotating condensates where they initiate  vortex lattice formation \cite{Sinha01, Parker06b}.  However, in the current study we will focus on stable excitations of non-rotating systems.

To find such eigenfunctions and eigenvalues we consider a polynomial ansatz for the perturbations in the coordinates $x,y$, and $z$, of a total degree $\nu$ \cite{Sinha01}, that is,
\begin{equation}
\delta n = \sum_{p,q,r} a_{pqr} x^p y^q z^r, \hspace{1cm} \delta \phase = \sum_{p,q,r} b_{pqr} x^p y^q z^r,
\label{eq:densityperturbation}
\end{equation}
where
\begin{equation}
\nu = \max_{\substack{a_{pqr}\neq 0\\b_{pqr} \neq 0}} \left\{ p + q + r \right\}.
\end{equation}
All operators in Eq.\ (\ref{pertop}), acting on such polynomials of degree $\nu$, result again in polynomials of the same order. For the operator $K$ this property might not be obvious, but a remarkable result known from $19^{th}$-century gravitational potential theory states that the integral in Eq.\ (\ref{Koperator}) evaluated for a polynomial density $\delta n$, yields another polynomial in $x, y$, and $z$. Its coefficients are given in terms of the integrals $\beta_{ijk}$ defined in Eq.\ (\ref{Betalmn}), and the exact expressions are presented in Appendix \ref{AppPotential}. The degree of the resulting polynomial is $\nu + 2$, and taking the derivative with respect to $z$ twice yields another polynomial of degree $\nu$ again. Thus, operator (\ref{pertop}) can be rewritten as a matrix mapping between scalar vectors of polynomial coefficients. Numerically finding the eigenvalues and eigenvectors of such a system is a simple task, which computational packages can typically perform.

We present only the lowest-lying shape oscillations corresponding to polynomial phase and density perturbations of degree $\nu=1$ and $\nu=2$.  These form the monopole, dipole, quadrupole and scissors modes.  These excitations are illustrated schematically in Fig.~\ref{Modes} and described below, where we state only the form of the density perturbation $\delta n$, since it can be shown that the corresponding phase perturbation $\delta S$ always contains the same monomial terms.  Note that $a$, $b$, $c$ and $d$ are real positive coefficients.
\begin{itemize}
\item {\bf Dipole modes $D_x$, $D_y$ and $D_z$}: A centre-of-mass motion along each trap axis \cite{NoteDipoleMode}.  The $D_x$ mode, for instance, is characterised by $\delta n = \pm ax$.
 \item {\bf Monopole mode $M$}: An in-phase oscillation of all radii with the form $\delta n=\pm a\pm(bx^2+cy^2+dz^2)$.
\item {\bf Quadrupole modes $Q_1^{xy}$, $Q_1^{xz}$ and $Q_1^{yz}$}: The $Q_1$ modes feature two radii oscillating in-phase with each other (denoted in superscripts) and out-of-phase with the remaining radius. For example, the $Q_1^{xy}$ mode is characterised by $\delta n=\pm a \pm (bx^2+cy^2-dz^2)$.
\item {\bf Quadrupole mode $Q_2$}:  This 2D mode is supported only in a plane where the trapping has circular symmetry.  For example, in the transverse plane of a cylindrically-symmetric system the transverse radii oscillate out-of-phase with each other, with no motion in $z$, according to $\delta n=\pm a(x\pm iy)^2$.

 \item {\bf Scissors modes $Sc_{xy}$, $Sc_{yz}$ and $Sc_{xz}$}: Shape preserving rotation of the BEC over a small angle in the $xy$, $xz$ and $yz$ plane, respectively.  The $Sc_{xy}$ mode is characterised by $\delta n=\pm a xy$.  Note that a scissors mode in a given plane requires that the condensate asymmetry in that plane is non-zero otherwise no cross-terms exist.  Furthermore, the amplitude of the cross-terms should remain smaller than the condensate/trap asymmetry otherwise the scissors mode turns into a quadrupole mode \cite{Guery99}. 
\end{itemize}
Note that, in order to confirm the dynamical stability of the solution, one must also check that positive eigenvalues do not exist.   We have performed this throughout this paper and consistently observe that when Im$(\lambda)\neq0$ that Re$(\lambda)=0$ and that when Im$(\lambda)=0$ that Re$(\lambda)\neq0$. 
It is also possible to determine excitation frequencies of higher order excitations of the BEC by including higher order monomial terms. Such modes, for example, play an important role in the dynamical instability of rotating systems \cite{Sinha01, Bijnen07}.

We would like to remind the reader that they can download the MATLAB program  \cite{FrequenciesProgram} used to perform the calculations described in this section. It includes an easy to use  graphical user interface.

\end{section}

\begin{section}{Excitations in a cylindrically-symmetric system \label{SecExcCyl}}
In this section we present the oscillation frequencies of the lowest lying stable excitations of a dipolar condensate in a cylindrically-symmetric trap.  Through specific examples we indicate how they behave with the key experimental parameters, namely the  dipolar interaction strength $\edd$ and trap ratio $\gamma$.  Note that we will discuss the scissors modes in more detail in Section \ref{SecScissors}.  Here we will just point out that two scissors modes exist, corresponding to $Sc_{xz}$ and $Sc_{yz}$, while the $Sc_{xy}$ mode is non-existant due to the cylindrical symmetry of the system. 

\subsection{Variation with dipolar interactions $\edd$ for $g>0$}

In Fig. \ref{FigCylFreq_gamma} we show how the collective mode frequencies vary with the dipolar interactions for the case of $g>0$.  Although it would seem experimentally relevant  to present these frequencies as a function of $\edd$, we plot them as a function of the aspect ratio $\kappa$ instead. We do this for the following two reasons: i) plotting the frequencies as a function of $\edd$ is problematic since two static solutions (metastable local minima and unstable saddle points) can exist for a given value of $\edd$; ii) in the critical region of collapse at the turning point from stable to unstable, the excitation frequencies vary rapidly as a function of $\edd$, but much more smoothly as a function of $\kappa$, and so it is easier to view the behavior  as a function of $\kappa$. 
For completeness we have included the corresponding plot of the frequencies, but as a function of $\edd$, in Appendix \ref{AppCollectiveepsilondd}. Also, analytic expressions for the frequencies  of the $M$ and $Q_{1}$ modes in a cylindrically symmetric dipolar BEC in the TF regime can be found in \cite{ODell04}. 

It is worth pointing out that the condensate shape accounts for a significant part of the physics of these systems, and so $\kappa$ is a good variable to work with. For example, in the problem of a rotating dipolar BEC, the critical rotation frequency at which a vortex becomes energetically favorable is exactly the same as that in a purely $s$-wave BEC providing one corrects for the change in the aspect ratio due to the dipolar interactions \cite{ODell07}. However, $\kappa$ alone does not contain all the physics. In the case of the calculation of the excitation frequencies this is clear from Eq.\ (\ref{pertop}) which depends upon both $(\nabla \cdot n_{0} \nabla) \delta S$ and $\edd K \delta n$. The former term has a direct dependence upon $\kappa$ via the equilibrium density profile $n_{0}(\rvec)$,   whereas the latter term does not.

We consider three values of trap ratio $\gamma$, which fall into three distinct regimes: (1) $\gamma < \gamma_{\rm crit}^-$, (2) $\gamma_{\rm crit}^- < \gamma < \gamma_{\rm crit}^+$ and (3) $\gamma_{\rm crit}^+ < \gamma$.  Recall that $\gamma_{\rm crit}^+ (\gamma_{\rm crit}^-)$ is the critical value above (below) which there exist stable solutions for $\edd \rightarrow +\infty (-\infty)$, see also Fig.\ \ref{FigStaticCyl1}.  In each case the aspect ratio of the {\em stable} solutions exists over a finite range $\kappa=[\kappa^-,\kappa^+]$. We will now discuss each regime in turn.

\begin{figure}[t]
\includegraphics[width=0.75\columnwidth]{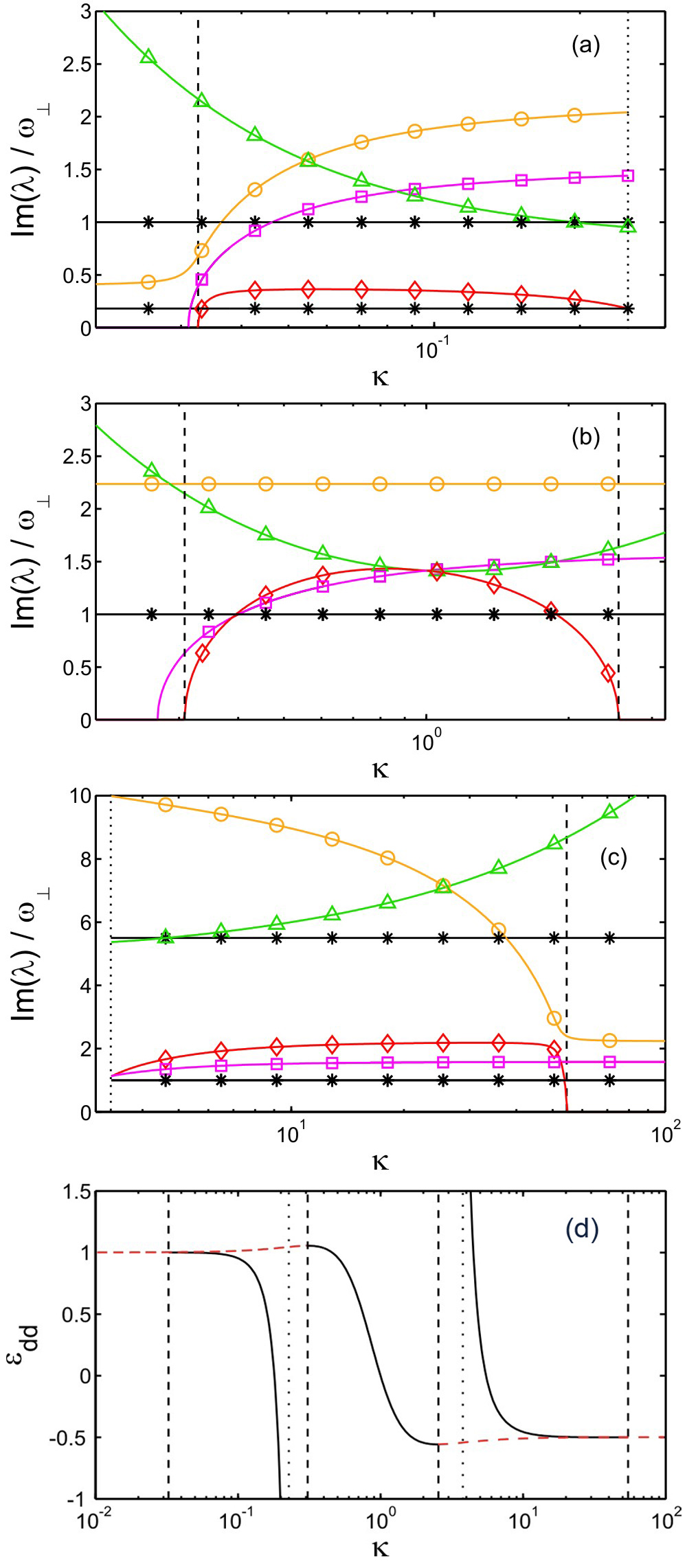}
\caption{(Color online) Excitation frequencies as a function of condensate aspect ratio $\kappa$ for a cylindrically-symmetric trap with aspect ratio (a) $\gamma = 0.18$, (b) $\gamma= 1 $ and (c) $\gamma = 5.5$. Shown are the results for the modes $M$ (orange, circles), $D$ (black, stars), $Q_1$ (red, diamonds), $Q_2$ (purple, squares) and $Sc_{xz} (= Sc_{yz})$ (green, triangles).  (d) Static solutions $\kappa$ for $\gamma=0.18$, $1$ and $5.5$. Vertical dashed lines mark the transition from stable to unstable for the static solution, and this coincides with the point at which one of the frequencies tends to zero. Vertical dotted lines mark the point at which the static solution ceases to exist altogether. \label{FigCylFreq_gamma}}
\end{figure}
\subsubsection{$\gamma < \gamma_{\rm crit}^-$}
In Fig.~\ref{FigCylFreq_gamma}(a) we present the excitation frequencies for $\gamma=0.18$ as a function of $\kappa$.  The corresponding static solutions are shown as the left hand curve in Fig.~\ref{FigCylFreq_gamma}(d) and confirm that the {\em stable} static solutions (solid black part of curve) exist only over a range of $\kappa=[\kappa^-,\kappa^+]$, with $\kappa^- \approx 0.03$ and $\kappa^+ \approx 0.25$ indicated by vertical lines (dashed and dotted, respectively).  For $\kappa>\kappa^+$, no static solutions exist and so the excitation frequencies are not plotted beyond this point [dotted vertical line in Fig.~\ref{FigCylFreq_gamma}(a) and most left hand dotted vertical line in \ref{FigCylFreq_gamma}(d)]. For $\kappa<\kappa^-$, the static solution is no longer a local energy minimum but becomes instead a saddle point/maximum that is unstable to collapse [transition marked with dashed, vertical line in Fig.~\ref{FigCylFreq_gamma}(a) and most left hand dashed vertical line in \ref{FigCylFreq_gamma}(d)].  Although this solution is not stable we can still determine its excitation spectrum.  Crucially, this will reveal which modes are responsible for collapse and which remain stable throughout.

Three dipole modes (stars) exist.  Dipole modes, in general, are decoupled from the internal dynamics of the condensate \cite{Stringari} and are determined by the trap frequencies $\omx, \omy$, and $\omz$. This provides an important check on our code. For the cylindrically symmetric case, $\omx = \omy = \operp$, and hence only two distinct dipole modes are visible.  For $\kappa<\kappa^-$ the dipole frequencies remain constant, indicating the dynamical stability of this mode.  

In general, the remaining modes vary with the dipolar interactions.  Perhaps the key mode here is the quadrupole $Q_1$ mode (diamonds).  At the point of collapse the $Q_1$ frequency decreases to zero.  This is connected to the dynamical instability of this mode since Re$(\lambda)> 0$ for $\kappa<\kappa^-$.  The physical interpretation of this is that the $Q_1$ mode, which comprises of an anisotropic oscillation in which the condensate periodically elongates and then flattens, mediates the collapse of the condensate into an infinitely narrow cigar-shaped BEC.  In the energy landscape picture, this occurs because the barrier between the local energy minimum and the collapsed $R_{x,y}=0$ state disappears for $\kappa<\kappa^-$.  Note that the link between collapse and the decrease of the quadrupole mode frequency to zero has been made in Ref.~\cite{Goral02}.  The $Q_2$ quadrupole mode (squares) decreases to zero, and becomes dynamically unstable, after one passes into the unstable regime as indicated in Fig.~\ref{FigCylFreq_gamma}(d).  The monopole $M$ mode (circles) remains stable for $\kappa<\kappa^-$ and increases with $\kappa$ above this point.

\subsubsection{$\gamma_{\rm crit}^- < \gamma < \gamma_{\rm crit}^+$}
In Fig.\ \ref{FigCylFreq_gamma}(b) we present the excitation frequencies for $\gamma=1$ as a function of $\edd$.  Since $\gamma_{\rm crit}^- < \gamma < \gamma_{\rm crit}^+$, the solutions exist over a finite range of $\edd$.  In terms of $\kappa$, collapse occurs at both limits of its range, i.e., for $\kappa<\kappa^-$ and $\kappa>\kappa^+$, where $\kappa^- \approx 0.3$ and $\kappa^+ \approx 2.5$ (dashed vertical lines in Fig.~\ref{FigCylFreq_gamma}(b) and (d)).  
 
Since the trap is spherically-symmetric, the dipole modes (stars) all have identical frequency, i.e. $\omega_\perp$.  The $Q_1$ quadrupole frequency (diamonds) decreases to zero at both points of collapse, $\kappa^-$ and $\kappa^+$.  In the former case, this corresponds to the anisotropic collapse into an infinitely narrow BEC, while in the latter case, collapse occurs into an infinitely flattened BEC.  In the low $\kappa$ regime, the $Q_2$ quadrupole mode (squares) becomes unstable just past the point of collapse, but shows no instability in the opposite limit for $\kappa>\kappa^+$. 

It is interesting to note that the monopole mode (circles) shows no dependence on $\kappa$ and therefore the dipolar interactions, in agreement with \cite{ODell04}. Additionally, we find that the aspect ratio of the density perturbation remains fixed at precisely $1$ for all values of the condensate aspect ratio $\kappa$. These observations are specific to the case of $\gamma=1$. 

\subsubsection{$\gamma>\gamma_{\rm crit}^+$}
In Fig.\ \ref{FigCylFreq_gamma}(c) we plot the excitation frequencies for $\gamma=5.5$.  For $\kappa<\kappa^-$, no static solutions exist, and for $\kappa>\kappa^+$, no {\em stable} solutions exist.  Here $\kappa^- \approx 3.3$ and $\kappa^+ \approx 54$ (dotted and dashed vertical lines, respectively, in Fig.~\ref{FigCylFreq_gamma}(c) and (d)).

Again, the dipole modes are constant, while the remaining modes vary with dipolar interactions.  Apart from the quadrupole $Q_1$ mode, all modes are stable past the point of collapse, including the $Q_2$ quadrupole mode.  The $Q_1$ mode decreases to zero at the point when the condensate collapses to an infinitely flattened pancake BEC, which is again consistent with this mode mediating the anisotropic collapse.

\subsection{Variation with dipolar interactions $\edd$ for $g<0$}\label{SubSecNegg}
We now consider the analogous case but with $g<0$.  As shown in Section \ref{SubSecCyl2} stable solutions only exist for $\gamma>\gamma^+_{\rm crit}=5.17$ and $\gamma<\gamma^-_{\rm crit}=0.19$, with no  stable solutions existing in the range $\gamma^-_{\rm crit}<\gamma<\gamma^+_{\rm crit}$.  Hence we will only consider the two regimes of (1) $\gamma<\gamma^-_{\rm crit}$ and (2) $\gamma>\gamma^+_{\rm crit}$.  

\subsubsection{$\gamma< \gamma_{\rm crit}^-$}
In Fig.\ \ref{FigCylFreq_gamma2}(a) we present the excitation frequencies in a highly elongated trap $\gamma=0.18$.  Stable static solutions exist only for $\kappa^-<\kappa<\kappa^+$ where $\kappa^- \approx 0.25$ and $\kappa^+ \approx 0.29$.  In this regime we find that all collective frequencies are purely imaginary and finite, and therefore stable.  At the critical point for collapse $\kappa \approx 0.29$ the $Q_1$ mode frequency passes through zero and becomes purely real, signifying its dynamical instability.  This shows that, as for $g>0$, the $Q_1$ mode mediates collapse and therefore collapse proceeds in a highly anisotropic manner due to the anisotropic character of the dipolar interactions. The remaining modes do not become dynamically unstable past the critical point, and only vary weakly over the range of $\kappa$ shown. 
It should also be remarked that higher order modes with polynomial degree $\nu > 2$ also become unstable within the range $\kappa^-<\kappa<\kappa^+$ where no stable parabolic solutions lie, further highlighting the \textit{meta}stability of the $g<0$ states and confirming the relevance of the predictions made by the uniform-density Bogoliubov spectrum (\ref{eq:bogdispersion}) for a system in the TF regime.

\subsubsection{$\gamma> \gamma_{\rm crit}^+$}
Figure \ref{FigCylFreq_gamma2}(b) shows the mode frequencies in a highly flattened trap $\gamma=5.5$, for which stable static solutions exist only in the regime $\kappa^-<\kappa<\kappa^+$ where $\kappa^-\approx 2.7$ and $\kappa^+ \approx 3.3$.  Similarly, at the point of collapse $\kappa\approx 2.7$ the $Q_1$ mode has zero frequency and is dynamically unstable.  Well below the critical point the $Q_2$ mode frequency also becomes zero and dynamically unstable.  

\begin{figure}[t]

\includegraphics[width=0.85\columnwidth]{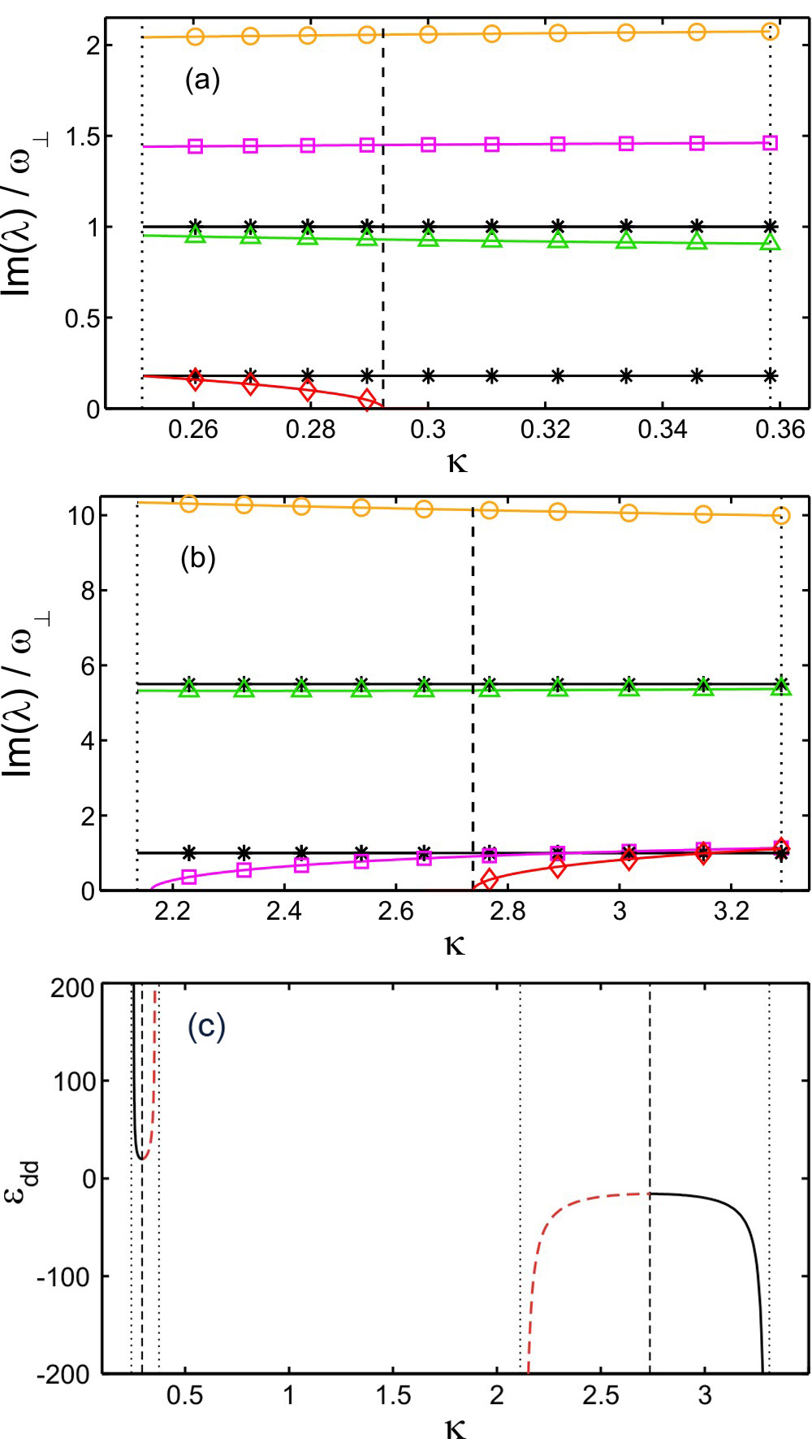}
\caption{(Color online) Excitation frequencies as a function of condensate aspect ratio $\kappa$ for a $g<0$ cylindrically-symmetric trap with aspect ratio (a) $\gamma = 0.18$ and (b) $\gamma = 5.5$, with corresponding static solutions shown in figure (c). Included are the results for the modes $M$ (orange, circles), $D$ (black, stars), $Q_1$ (red, diamonds), $Q_2$ (purple, squares) and $Sc$ (green, triangles). Dashed vertical lines indicate the critical point at which the stable static solutions turn into unstable ones, dotted vertical lines indicate endpoints of branches where static solutions cease to exist. \label{FigCylFreq_gamma2}}
\end{figure}

\subsection{Variation with trap ratio $\gamma$}
Having illustrated in the previous section how the excitation frequencies behave for $g<0$, from now on we will limit ourselves to the case of $g>0$.  In Fig.~\ref{FigCylFreq_edd} we plot the excitation frequencies as a function of $\gamma$ for various values of $\edd$.    A common feature is that the dipole frequencies scale with their corresponding trap frequencies, such that $\omega_{D_x}=\omega_{D_y}=\omega_\perp$ and $\omega_{D_z}=\gamma\omega_\perp$.  We now consider the three regimes of zero, negative and positive $\edd$.
\begin{figure}
\includegraphics[width=0.85\columnwidth]{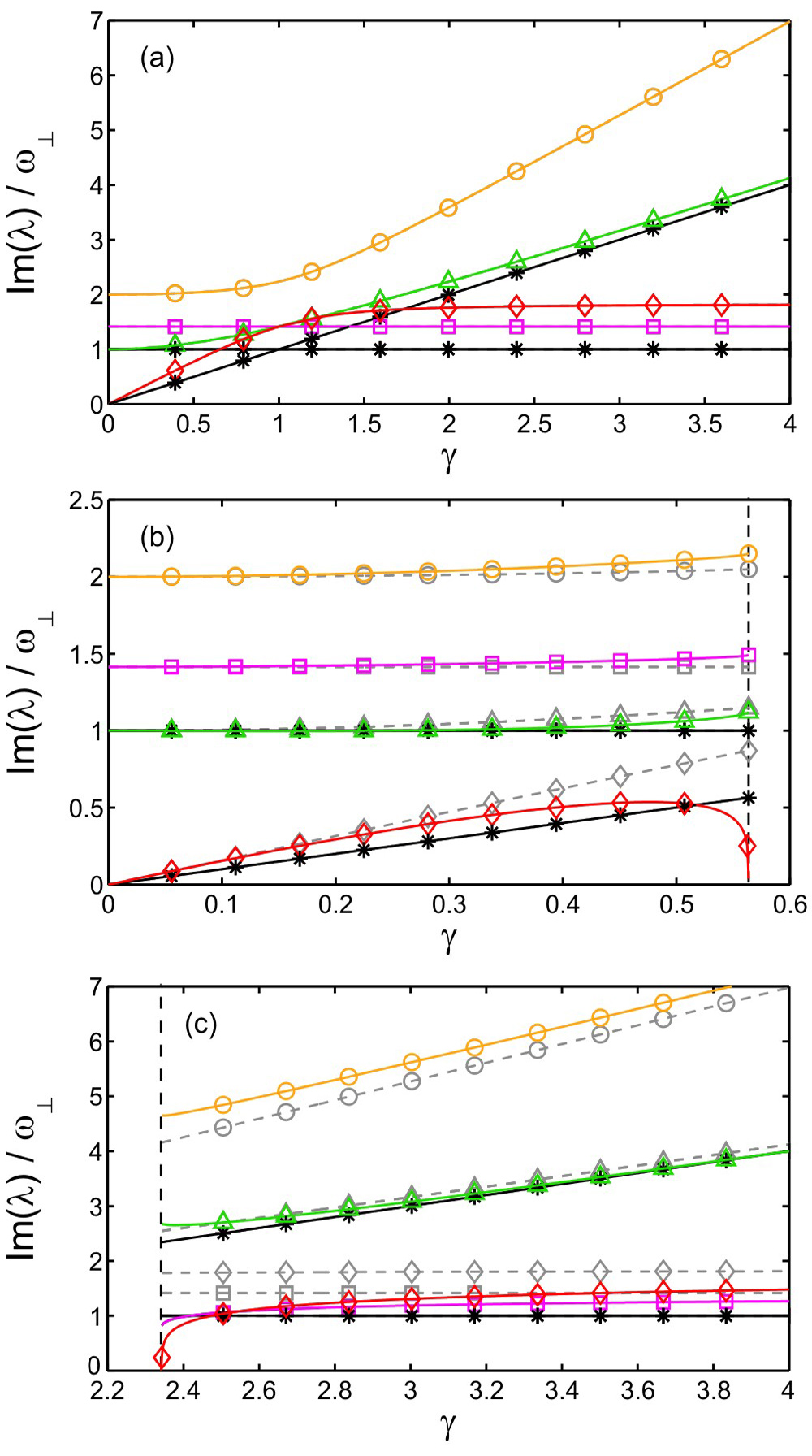}
\caption{(Color online) Excitation frequencies in a cylindrically-symmetric trap as a function of the trap aspect ratio $\gamma$ for (a) $\edd=0$, (b) $\edd=-0.75$ and (c) $\edd=1.5$.  Shown are the results for the modes $M$ (orange, circles), $D$ (black, stars), $Q_1$ (red, diamonds), $Q_2$ (purple, squares) and $Sc$ (green, triangles). In figures (b) and (c) the frequencies for $\edd = 0$ are included as dashed, gray lines.\label{FigCylFreq_edd}}
\end{figure}

\subsubsection{$\edd=0$}
For $\edd=0$ stable solutions exist for all $\gamma$ and the corresponding mode frequencies are plotted in Fig.~\ref{FigCylFreq_edd}(a).  Our results agree with previous studies of non-dipolar BECs where analytic expressions for the mode frequencies can be obtained, see, e.g., \cite{PethickSmith} and \cite{Stringari}.  The $Q_2$ quadrupole mode has fixed frequency $\omega_{Q_2}=\sqrt{2}\omega_\perp$.  The scissors mode frequency corresponds to $\omega_{Sc_{xz}}=\omega_{Sc_{yz}}=\sqrt{1+\gamma^2}\omega_{\perp}$, and the remaining modes obey the equation \cite{Stringari},
\begin{equation}
 \omega^2=\omega_\perp^2\left( 2+\frac{3}{2}\gamma^2\pm\frac{1}{2}\sqrt{16-16\gamma^2+9\gamma^4} \right),
\end{equation}
where the ``+'' and ``-'' solutions correspond to $\omega_M$ and $\omega_{Q_1}$, respectively.

\subsubsection{$\edd<0$}
For $\edd=-0.75$ (Fig.~\ref{FigCylFreq_edd}(b)) stable solutions, and collective modes, exist up to a critical trap ratio $\gamma^{\rm max}\approx 0.56$.  Beyond that the attractive nature of side-by-side dipoles (recall $\Cdd<0$) makes the system unstable to collapse.  

For all of the modes except the $ Q_1$ quadrupole mode we see the same qualitative behaviour as for the non-dipolar case (grey lines) with the modes extending right up to the point of collapse with no qualitative distinction from the non-dipolar case.  The $Q_1$ quadrupole mode, on the other hand, initially increases with $\gamma$, like the non-dipolar case, but as it approaches the point of collapse, it rapidly decreases towards zero.  Above $\gamma^{\rm max}$, the $Q_1$ mode is dynamically unstable.

\subsubsection{$\edd>0$}
For $\edd=1.5$ (Fig.~\ref{FigCylFreq_edd}(c)) stable solutions exist only above a lower critical trap ratio $\gamma^{\rm min}\approx 2.3$.  For $\gamma<\gamma^{\rm min}$ the attraction of the end-to-end dipoles becomes dominant and induces collapse.  Indeed, we find that the frequency of the $Q_1$ mode passes through zero and is dynamically unstable for $\gamma<\gamma^{\rm min}$.  Above this, the $Q_1$ and $Q_2$ frequencies increase towards the limiting values of the non-dipolar frequencies of $1.82 \omega_\perp$ and $\sqrt{2}\omega_\perp$ because in a very pancake-shaped trap the atoms cannot sample the anisotropy of the interactions.  The remaining modes behave qualitatively like the non-dipolar modes for $\gamma>\gamma^{\rm min}$.

\end{section}

 \begin{section}{Non-cylindrically-symmetric systems and relevance to rotating-trap systems\label{SecExcNonCyl}}
 In this section we will apply our approach to the most general case of non-cylindrically-symmetric systems.  An important experimental scenario where this occurs is when condensates are rotated in elliptical harmonic traps.  This has provided a robust method for generating vortices and vortex lattices in condensates (see Ref.~\cite{emergent} for a review). Whilst the trap ellipticity in the $x-y$ plane is typically small (in most experiments it is of the order of a few percent), the rotation accentuates the ellipticity induced in the condensate.  Indeed, one can derive effective harmonic trap frequencies for the condensate which show that the effective ellipticity can be orders of magnitude greater than the static ellipticity \cite{Recati01,Sinha01,Bijnen07}.  
 
The $Q_2$ mode can be pictured as a surface wave traveling around the edge of the condensate. It has a similar shape to the rotating elliptical deformation of the trap, and when the trap is rotated at frequencies close to that of the $Q_2$ mode then even a perturbatively small trap deformation strongly couples to this mode. When viewed from the frame of reference rotating with the trap, the excitation of the $Q_2$ mode  appears as a bifurcation of the stationary condensate into a new stationary state which mixes in some of the $Q_2$ mode and the condensate therefore develops an elliptical shape in the $x-y$ plane. For some ranges of rotation speeds this new stationary state is in turn dynamically unstable to the excitation of higher order modes  \cite{Sinha01,Bijnen07,Bijnen08a}. This dynamical instability disrupts the condensate and is the first step in the process by which vortices enter.
Although this process is complex, the dynamical instability that initiates it is accurately described within the TF approximation because the modes which are initially excited are of sufficiently long wavelength.  The predictions obtained within the TF approximation are in excellent agreement with both experiments \cite{Madison01} and numerical simulations of the GPE \cite{Parker06b,Corro07}.  Although we will not specifically consider rotation further here, our methodology can be easily extended to this scenario \cite{Bijnen07}.

  \begin{figure}[t]
 \includegraphics[width=0.83\columnwidth]{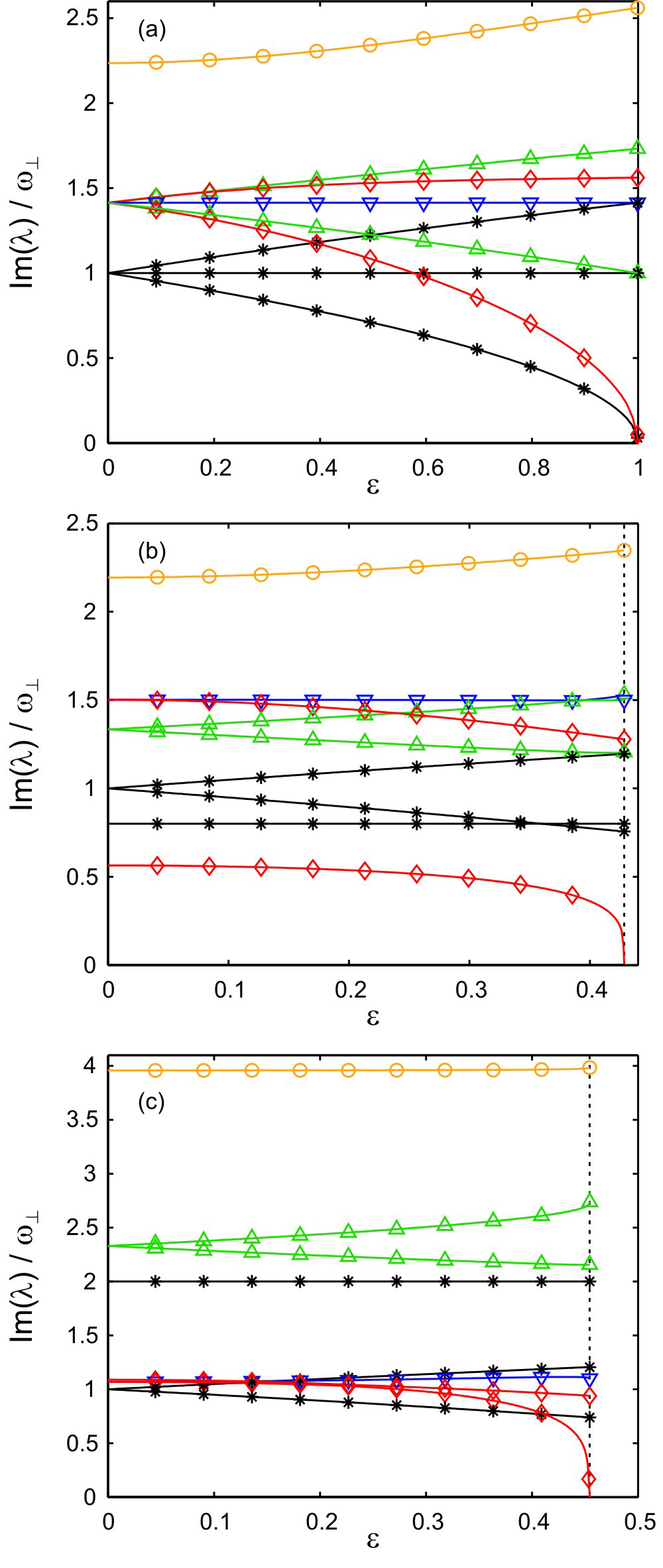}
 \caption{(Color online) Excitation frequencies in a non-cylindrically-symmetric trap as a function of the trap ellipticity $\epsilon$ in the $x-y$ plane for the cases of (a) $\edd=0$ and $\gamma=1$, (b) $\edd=-0.6$ and $\gamma=0.8$ and (c) $\edd=1.25$ and $\gamma=2$.  Shown are the modes $D$ (black, stars), $M$ (orange, circles), $Q_1$ (red, diamonds), $Sc_{xy}$ (blue, triangles pointing down), $Sc_{yz}$ (green, triangles pointing up, upper branch) and $Sc_{xz}$ (green, triangles pointing up, lower branch).  \label{FigAnisotropicFreq}}
 \end{figure}

As in Section \ref{SubSecAnisotropic}, we consider finite trap ellipticity $\epsilon$ in the $x-y$ plane.  In Fig.~\ref{FigAnisotropicFreq} we present the mode frequencies as a function of ellipticity $\epsilon$ for three different examples. There are some important generic differences to the cylindrical case.  Due to the complete anisotropy of the trapping potential the dipole mode frequencies (stars) all differ, and are equal to the corresponding trap frequencies $\omega_x=\sqrt{1-\epsilon}\omega_\perp$, $\omega_y=\sqrt{1+\epsilon}\omega_\perp$ and $\omega_z=\gamma\omega_\perp$.  The monopole mode is present (circles) and its frequency increases with $\epsilon$. Strictly speaking the $Q_2$ mode is no longer present due to the breakdown of cylindrical symmetry.  Instead we find a new $Q_1$ mode appearing (upper diamonds) which corresponds to the $Q_1^{yz}$ mode for $\edd>0$ and the $Q_1^{xz}$ mode for $\edd \leq 0$. The usual quadrupole mode $Q_1^{xy}$ is also present. The reader is reminded that the superscript in the $Q_1$ mode notation refers to the \textit{in-phase} radii, the remaining radius oscillates out of phases with the other two. Although there are actually three permutations of $Q_1$, only two appear for any given value of $\edd$ since linear combinations of these and the monopole mode can form the remaining $Q_1$ mode.

We will now consider the specific features for the cases presented in Fig.~\ref{FigAnisotropicFreq}.
For $\edd=0$ and $\gamma=1$ (Fig.~\ref{FigAnisotropicFreq}(a)), the solutions are stable right up to $\epsilon=1$.  At this limit the {\it x}-direction becomes untrapped and this causes the system to become unstable with respect to the dipole $D_x$ mode, as well the $Q_1^{xy}$ mode which can now expand freely along the {\it x}-axis.  For $\edd=-0.6$ and $\gamma=0.8$ (Fig.~\ref{FigAnisotropicFreq}(b)) the solutions become unstable to collapse at $\epsilon \approx 0.425$.  Only the lower $Q_1^{xy}$ mode becomes dynamically unstable at this point, indicating that it is the mode responsible for collapse, which is towards a pancake shaped system.  For $\edd=1.25$ (Fig.~\ref{FigAnisotropicFreq}(c)) the solutions become unstable to collapse at $\epsilon \approx 0.45$.  We again observe that the same $Q_1$ mode  mediates the collapse, only this time the collapse is towards a cigar shaped system. The other modes remain stable.

 \end{section}

\begin{section}{Scissors modes \label{SecScissors}}
A fundamental question concerning ultracold dipolar Bose gases  is the nature of their quantum state in situations when the attractive portion of their interactions becomes important, such as in cigar-shaped systems aligned along the external polarizing field. Some time ago \cite{Huang87,Nozieres95} it was noticed that, due to exchange effects, repulsive interactions favor simple Bose-Einstein condensation (macroscopic occupation of a single quantum state) over fragmented Bose-Einstein condensation (macroscopic occupation of two or more quantum states) \cite{Leggett01}. The converse is true in the presence of attractive interactions. Fragmentation can therefore be potentially studied in attractive $s$-wave condensates which are stabilized by their zero-point energy. However, the consensus seems to be that in those systems mechanical collapse of the BEC occurs before significant fragmentation \cite{Mueller06}. Dipolar interactions, on the other hand, are partially attractive and partially repulsive, the net balance being tunable via the shape of the atomic cloud. A comprehensive investigation of fragmentation in dipolar BECs is beyond the scope of the current paper, but below we take a step in this direction by calculating the properties of scissors modes of a dipolar BEC.

Experimentally, one of the simplest indicators of whether or not an atomic cloud is Bose condensed is to examine the momentum distribution following free expansion after the trap is turned off \cite{BEC1995}. For example, according to the equipartition theorem, a gas at thermal equilibrium will expand isotropically even if the trap was anisotropic. This is not true for a BEC which, due to its zero-point energy, expands most rapidly in the direction which was most tightly confined. However, for dipolar BECs the situation is complicated by the anisotropy of the long-range interactions which continue to act at some level even as the gas expands \cite{Stuhler05,Lahaye07}. Quantized vortices are  another ``smoking gun'' indicating the presence of a BEC, but these are not easy to controllably generate in the cigar-shaped systems which would be of primary interest (although they might be useful in cases where $\Cdd < 0$, for which pancake-shaped BECs have dominant attractive interactions). Furthermore, in cigar-shaped systems with $\Cdd > 0$, the rotation speed at which a vortex becomes energetically favorable diverges as $\edd$ increases \cite{ODell07}. Scissors modes, on the other hand, offer an alternative vehicle for the investigation of superfluidity in dipolar systems which does not suffer from the difficulties mentioned above.

A detailed account of the scissors mode in a pure $s$-wave BEC can be found in \cite{Guery99}.  The scissors mode of a trapped atomic cloud (thermal or Bose condensed) is excited by suddenly rotating the anisotropic trapping potential over a small angle. Consequently, the atomic cloud will experience a restoring force exerted by the trap, and provided the angle of rotation is small, it will exhibit a shape preserving oscillation around the new equilibrium position. The exact response of the atomic cloud to the torque of the rotated trapping potential depends strongly on the moment of inertia of the cloud. Since a superfluid is restricted to irrotational flow, it will have a significantly different moment of inertia compared to a thermal cloud. In particular, when the trap anisotropy vanishes the moment of inertia of a superfluid also vanishes, whereas in a thermal cloud this is not the case. The superfluid scissors mode frequency will consequently approach a finite value, whereas in a thermal cloud it will vanish as the trap anisotropy approaches zero \cite{Guery99}. A measurement of the scissors mode frequency therefore constitutes a direct test for superfluidity \cite{Guery99,Stringari}, as has been verified experimentally for non-dipolar BECs \cite{Marago00,Cozzini03}.  

In the following, we will consider the scissors mode to be excited by rotating the trapping potential as well as the external aligning field of the dipoles simultaneously and abruptly through a small angle, such that the condensate suddenly finds itself in a rotationally displaced configuration. Three scissors modes now appear due to the three distinct permutations of this mode, namely $Sc_{xy}$ (triangles pointing down in Fig. \ref{FigAnisotropicFreq}), $Sc_{yz}$ (triangles pointing up), and $Sc_{xz}$ (triangles pointing up).  Clearly, from Fig. \ref{FigAnisotropicFreq}, the oscillation frequencies of the scissors modes are affected by the dipolar interactions. The effect of the dipolar interactions is two-fold. Firstly, since the dipolar interactions change the aspect ratio of the condensate, both the moment of inertia of the condensate and the torque from the trapping potential acting on it will be altered, which consequently will alter the oscillation frequency. Secondly, for the $Sc_{xz}, Sc_{yz}$ modes there is an additional force present which is related to the relative position of the dipoles. This effect is easiest understood when considering a cigar shaped condensate. When such a condensate is rotated with respect to the aligning field, the dipoles are on average slightly more side-by-side than in the equilibrium situation. 
\begin{figure}
\includegraphics[width=0.85\columnwidth]{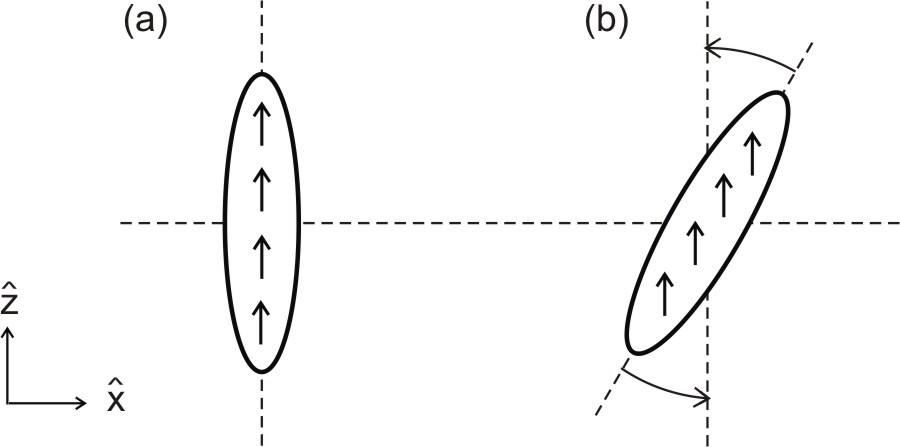} 
\caption{Schematic illustration of the dipolar restoring force for the $Sc_{xz}$ mode. When the condensate is rotated with respect to the dipole alignment axis $\hat{z}$ as in situation (b), the dipoles will on average be more side-by-side than in the aligned case, situation (a). Since this is an energetically unfavourable configuration compared to the aligned case, there will consequently be a dipolar restoring force present in (b) trying to re-align the condensate, illustrated by arrows.\label{FigDipolarRestoringForce}}
\end{figure}
As a result, there will be a dipolar restoring force trying to re-align the dipoles, which in turn is expected to affect the scissors mode frequencies. Figure \ref{FigDipolarRestoringForce}  schematically illustrates this process for the $Sc_{xz}$ mode. For a pancake shaped condensate the effect is opposite. Since the dipolar interaction potential is rotationally invariant in the $xy$ plane, the dipolar restoring force is absent for the $Sc_{xy}$ mode. 

Explicit expressions for the scissors frequencies can be obtained by performing the procedure outlined in section \ref{SecExcSpec} analytically, rather than numerically. We start with the frequency $\omega_{sxy}$ of the $Sc_{xy}$ mode, in which case we only expect an influence of dipolar interactions through changes in the geometry, and find 
\begin{equation}\label{Scxy}
\omega_{sxy}^2 = 2 \operp^2 \epsilon \left( \frac{\kappa_x^2 - \kappa_y^2}{\kappa_y^2 + \kappa_x^2} \right)^{-1},
\end{equation}
where it should be noted that the quantity in brackets is precisely the ellipticity of the condensate. The $Sc_{xy}$ frequency does not depend explicitly on the strength of the dipolar interactions $\edd$, but merely on the condensate ellipticity, which is an indication of the absence of a dipolar restoring force as discussed above. The condensate ellipticity turns out to be approximately proportional to the trap ellipticity, where the constant of proportionality  is dependent on the dipolar interaction strength $\edd$ and axial trapping strength $\gamma$. As a result, the $Sc_{xy}$ scissors frequencies shown in Fig. \ref{FigAnisotropicFreq} are (almost) independent of the trap ellipticity for fixed values of $\edd$ and $\gamma$. Figure \ref{FigScxy}(a) shows the $Sc_{xy}$ frequency as a function of the axial trapping strength $\gamma$, for various dipolar interaction strengths $\edd$. 
\begin{figure}
\includegraphics[width=0.85\columnwidth]{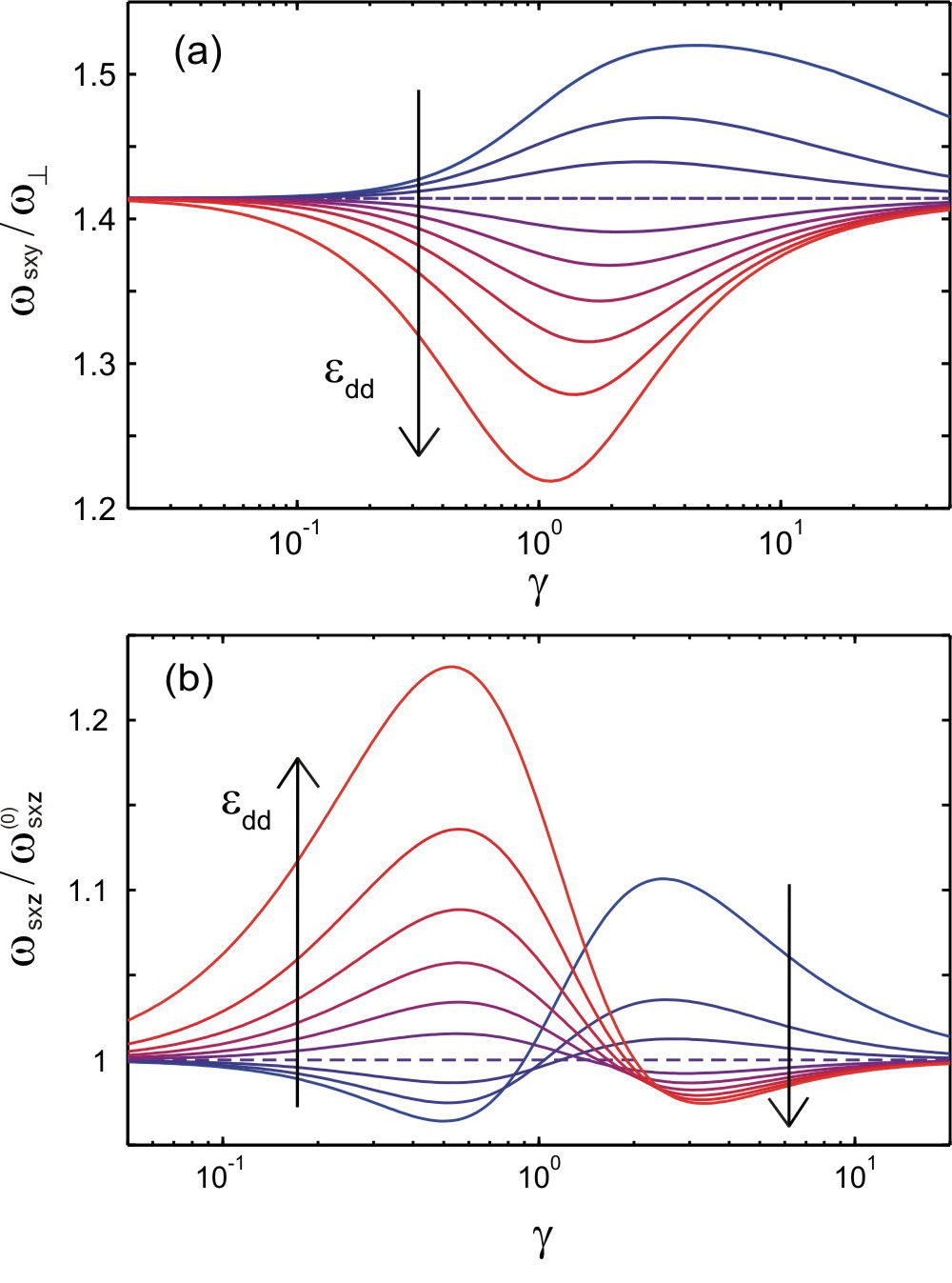} 
\caption{(Color online) Scissors frequencies as a function of the axial trapping strength $\gamma$ for various values of $\edd$, with $-0.45 \leq \edd \leq 0.9$ and increasing in the direction of the arrow in steps of $0.15$. The dashed line indicates $\edd = 0$. (a) Frequency of $Sc_{xy}$ mode for fixed trap ellipticity of $\epsilon = 0.1$. For very prolate $(\gamma \ll 1)$ or very oblate $(\gamma \gg 1)$ systems, the dipolar interactions renormalize into the $s$-wave interactions and $\omega_{sxy}$ returns to the non-dipolar value. (b) Frequency $\omega_{sxz}$ of the $Sc_{xz}$ mode as a function of $\gamma$ for a cylindrically symmetric trap, scaled to the non-dipolar frequency $\omega_{sxz}^{(0)}$.\label{FigScxy}}
\end{figure}
In the presence of dipolar interactions, the condensate ellipticity deviates from the trap ellipticity $\epsilon$ (see Section \ref{SubSecAnisotropic}), and hence the $Sc_{xy}$ frequency also changes when dipolar interactions are switched on. In the absence of dipolar interactions [dashed line in Fig. \ref{FigScxy}(a)], the trap and condensate ellipticity are equal and the $Sc_{xy}$ frequency is independent of the condensate size, trap ellipticity, as well as the $s$-wave interaction strength \cite{Guery99}. For very prolate ($\gamma \ll 1$) and very oblate ($\gamma \gg 1$) traps, the dipolar interactions become either mainly attractive or mainly repulsive and lose their anisotropic character. The dipolar potential becomes contactlike and can be renormalized into the $s$-wave interactions (see Section \ref{SubSecCyl2} or \cite{Parker08}), which do not influence the scissors mode frequency. This effect is visible in Fig. \ref{FigScxy}(a) in the form of the scissors frequency returning to the non-dipolar value for extremal values of $\gamma$. Finally, we would like to point out a remarkable similarity between the scissors frequencies shown in Fig. \ref{FigScxy}(a), and the trap rotation frequencies at which the static solution diagram of a rotating dipolar BEC shows a bifurcation point, as investigated in reference \cite{Bijnen08a} (see figure 1(b) therein). For all values of $\gamma$ and $\edd$ the scissors frequency is precisely twice the bifurcation frequency. Presumably, the underlying connection is the fact that the scissors mode $Sc_{xy}$ has the same superfluid field as a stationary state of a BEC in a rotating trap. However, a deeper investigation into the exact nature of the relationship is beyond the scope of this paper.

Turning our attention to the $Sc_{xz}$ and $Sc_{yz}$ frequencies, the analytical calculation yields
\begin{equation}\label{Scxz}
\frac{\omega_{sxz}^2}{\omega_z^2} = \left(\frac{1}{\kappa_x^2} + \frac{1}{\kappa_y^2}\right) \frac{1 - \edd\left(1 - \frac{9}{2} \kappa_x^3 \kappa_y \beta_{102}\right)}{1 - \edd\left(1 - \frac{9}{2} \kappa_x \kappa_y \beta_{002}\right)},
\end{equation}
\begin{equation}\label{Scyz}
\frac{\omega_{syz}^2}{\omega_z^2} = \left(\frac{1}{\kappa_x^2} + \frac{1}{\kappa_y^2}\right) \frac{1 - \edd\left(1 - \frac{9}{2} \kappa_x \kappa_y^3 \beta_{012}\right)}{1 - \edd\left(1 - \frac{9}{2} \kappa_x \kappa_y \beta_{002}\right)}.
\end{equation}
Here, the quantity $\edd$ appears explicitly and as such the frequencies depend directly on the strength of the dipolar interactions, an effect we attribute to the dipolar restoring force. Figure \ref{FigScxy}(b) shows the above frequencies for a cylindrically symmetric trap and as a function of the axial trapping strength $\gamma$, for various values of $\edd$. There are two distinct effects to be noted. Firstly, when $\edd > 0$ ($\edd < 0$) the scissors frequencies go up (down) for cigar shaped systems and down (up) for pancake shaped systems. This behaviour is consistent with what one would expect in the presence of a dipolar restoring force. Secondly, for $\gamma \ll 1$ and $\gamma \gg 1$ we see that the $\omega_{sxz}$ frequency approaches the non-dipolar value again. For the $\omega_{sxy}$ frequency this effect could be explained solely by the fact that for such values of $\gamma$ the condensate aspect ratios return to the non-dipolar values. However, for the $Sc_{xz}$ and $Sc_{yz}$ modes we have to account for the apparent vanishing of the dipolar restoring force as well. To see why it plays no part here, we have to analyze the expectation values of the quantity $R = \sqrt{x^2 + y^2 + z^2}$. For $\gamma \ll 1$ we have $\langle R \rangle \simeq \langle |z| \rangle \rightarrow \infty$, and for $\gamma \gg 1$ we have $\langle R \rangle \simeq \langle \sqrt{x^2 + y^2} \rangle \rightarrow \infty$. Although in both cases the torque exerted by the dipolar restoring force is proportional to $\langle R \rangle$ and in principle approaches infinity, it vanishes relative to the other two quantities contributing to the scissors frequencies, namely the moment of inertia of the condensate and the torque exerted by the trap, which both scale as $\langle R^2 \rangle$ \cite{Zambelli01}. In Fig. \ref{FigScxy}(b) this behaviour can be observed for the extremal values of $\gamma$, where the scissors frequencies approach that of the non-dipolar case.

\end{section}

\begin{section}{Conclusions}\label{SecConclusions}

In this paper we have performed an investigation into the static and dynamic states of trapped dipolar Bose-Einstein condensates in the Thomas-Fermi regime. We have extended our previous work in this area by examining new regimes  of dipolar and $s$-wave interactions (namely, positive and negative values of $\Cdd$ and $g$), non-cylindrically symmetric traps, and different classes of collective excitation, including the scissors modes. Our approach is based upon the analytic calculation of the non-local dipolar mean-field potential inside the condensate and allows us to calculate the potential due to an arbitrary polynomial density profile in an efficient manner.  Using this method,
 we have examined the stability of static states and collective excitations, including the behavior of the collective excitations as a function of the trap aspect ratio and ellipticity, and as a function of the relative strength of the dipolar and $s$-wave interactions. We consistently find that an instability of the $Q_{1}$ quadrupole mode mediates global collapse of a dipolar BEC whether $g>0$ or $g<0$. However, there are two critical trap ratios, $\gamma_{\rm crit}^+=5.17$ and $\gamma_{\rm crit}^-=0.19$,  beyond which the BEC is stable against scaling fluctuations (monopole and quadrupole excitations) even as the strength of the dipolar interaction overwhelms the $s$-wave one, i.e.\ when $\edd \rightarrow \pm \infty$. In the case of attractive $s$-wave interactions ($g<0$), where the dipolar interactions can stabilize an otherwise unstable condensate, the magnetostriction seems  to act counter-intuitively (see Figure \ref{FigStaticCyl2}), although upon closer examination the behavior can be explained by understanding how dipolar interactions behave in highly confined geometries.  
  
 We have paid special attention to the scissors modes because of their sensitivity to superfluidity, which we identify as an issue of particular interest in cigar-shaped dipolar condensates  due to the possibility of fragmentation when the attractive part of the dipolar interaction becomes significant. Our expressions for the frequencies of the scissors modes include a term due to a restoring force which is not present in the pure $s$-wave case, and which we identify as arising due to a long-range dipolar re-aligment force.
  
A freely available MATLAB implementation of the calculations outlined in this paper, including a graphical user interface, can be obtained online \cite{FrequenciesProgram}.

\end{section}

\begin{acknowledgments}
We acknowledge support from The Netherlands Organisation for Scientific Research (NWO) (R. M. W. van Bijnen and S. J. J. M. F. Kokkelmans), Canadian Commonwealth fellowship program (N. G. Parker), Australian Research Council (A. M. Martin) and Natural Sciences and Engineering Research Council of Canada (D. H. J. O'Dell). The authors also wish to thank T. Hortons for vital stimulation.  
\end{acknowledgments}

\appendix
\begin{section}{Collective modes frequencies as a function of $\edd$} \label{AppCollectiveepsilondd}
In Section \ref{SecExcSpec} we considered the effect of the dipolar interactions on the mode frequencies and plotted this as a function of $\kappa$ rather than $\edd$ to remove the problem of the static solutions being double-valued.  However, since $\edd$ is a more obvious experimental parameter, we have plotted the corresponding frequency plots of Fig. \ref{FigCylFreq_gamma}, but as a function of $\edd$ in Fig.~\ref{FigCylFreq_edd2}.  
\begin{figure}[t]
\includegraphics[width=0.8\columnwidth]{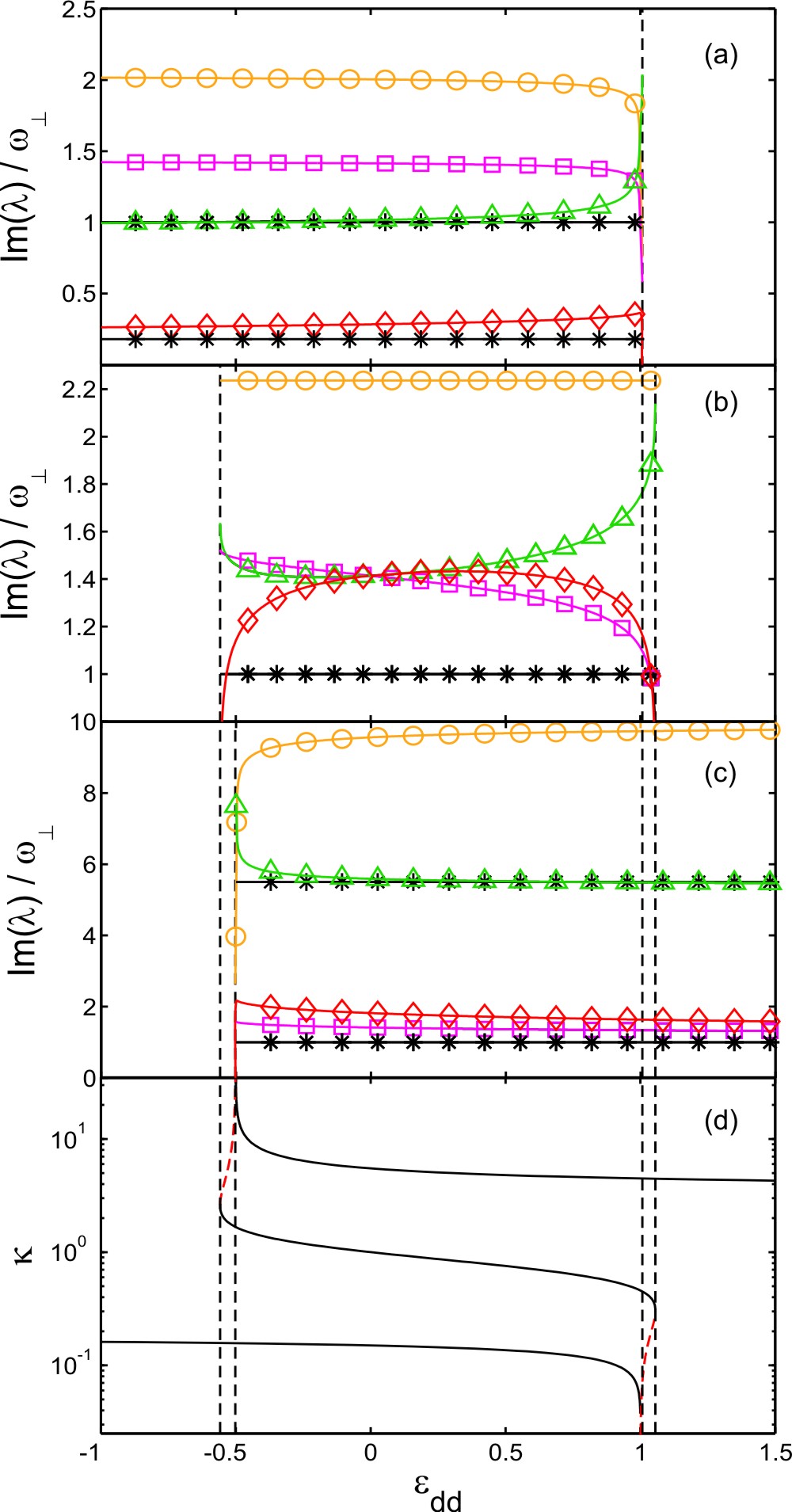}
\caption{Excitation frequencies as a function of $\edd$ for a cylindrically-symmetric trap with aspect ratio (a) $\gamma = 0.18$, (b) $\gamma= 1 $ and (c) $\gamma = 5.5$. Shown are the results for the dipole mode $D$ (black, stars), monopole $M$ (orange, circles), quadrupoles $Q_1$ (red, diamonds) and $Q_2$ (purple, squares), and scissors $Sc$ (green, triangles).  (d) Static solutions $\kappa$ for $\gamma=0.18$ (bottom curve), $1$ (center curve) and $5.5$ (top curve). Stable solutions are marked with a solid line, unstable solutions ar marked with a dashed (red) line. Dashed vertical lines mark the transition point from stable to unstable. \label{FigCylFreq_edd2}}
\end{figure}
\end{section}

\begin{section}{Calculating the dipolar potential inside a heterogenous ellipsoidal BEC\label{AppPotential}}
In this appendix we will concern ourselves with the calculation of integrals of the form
\begin{equation}\label{phirho}
\phi[\rho](\rvec) = \frac{1}{4\pi}\int \frac{\rho(\rvec')}{|\rvec' - \rvec|}\mathrm{d}x'\mathrm{d}y'\mathrm{d}z',
\end{equation}
where the domain of integration is a general ellipsoid with semi-axes $R_x, R_y, R_z$, and the point $\rvec = (x, y, z)$ is an internal point of the ellipsoid.  The square brackets indicate a functional dependence. In Eqs.\ (\ref{ElectrostaticPotential}) and (\ref{Koperator}) we need to evaluate this integral in order to obtain the fictitious electrostatic potential $\phi[\rho_{ijk}](\rvec)$  arising from a particle density of the form
\begin{equation}\label{rhoijk}
\rho_{ijk} = x^i y^j z^k,
\end{equation}
with $i,j,k$ nonnegative integers. By taking linear combinations of the general term $\rho_{ijk}$ we can calculate the internal dipolar potential created by an arbitrary density distribution because  Eq.\ (\ref{phirho}) defines a \emph{linear} integral operator acting upon $\rho(\rvec)$.

The physically relevant density distributions naturally fall into two classes: 
\begin{enumerate}
\item The inverted parabola $n(\rvec)$ given by Eq.\ (\ref{DensParabolic}) which corresponds to the static ground state of the BEC. 
\item The excitations $\delta n(\rvec)$ and $\delta S(\rvec)$ given by Eq.\ (\ref{eq:densityperturbation}) which can be written as linear combination of terms $x^{i}y^{j}z^{k}$.
\end{enumerate}
Actually, these two classes have some overlap because certain low lying excitations (monopole, quadrupole, and scissors modes), which would otherwise seem to fall into class 2, can be described in terms of the parabolic density profile of class 1 but with time-oscillating radii (in the case of monopole and quadrupole modes), or time-oscillating  symmetry axes (in the case of the scissors modes). In this appendix we give the general theory which works for all density distributions of the form (\ref{rhoijk}). In Appendix \ref{AppPotential2} we specialize to the parabolic density distributions of class 1 which is a particular case of the general theory and one is able to present the results in terms of well known special functions (the elliptic integrals).

In the context of calculating gravitational potentials in astrophysics one encounters exactly the same integrals as here and as such, the problem attracted considerable interest even in the $19^{th}$ century. Among others, significant contributions to the topic were made by MacLaurin, Ivory, Green, Poisson, Cayley, Ferrers, and Dyson. A detailed historical overview can be found, for instance, in references \cite{Chandrasekhar, Rahman01}. However, for our purposes the most important contribution came from N. M. Ferrers, who showed that the potential (\ref{phirho}) associated with the density (\ref{rhoijk}) evaluates exactly to a polynomial in the coordinates $x, y,$ and $z$ \cite{Ferrers1877}. As a matter of general interest we will outline the method employed by Ferrers to arrive at this remarkable result.

In his 1877 paper, Ferrers first shows how the internal (gravitational) potential of an ellipsoid with semi-axes $R_x, R_y,$ and $R_z$, with a density of the form
\begin{equation}\label{rhos}
\rho = \rho(s) = s^n, \hspace{1cm} n = 1, 2, 3, \ldots
\end{equation}
with
\begin{equation}\label{shomoeoid}
s = 1 - \frac{x^2}{R_x^2} - \frac{y^2}{R_y^2} - \frac{z^2}{R_z^2}
\end{equation}
can be calculated, using integration over so-called homoeoidal shells, to be
\begin{equation}\label{phisn}
\phi[s^n](x,y,z) = \frac{R_x R_y R_z}{4(n + 1)}\int_0^{\infty}Q^{n+1}\frac{\mathrm{d}\sigma}{\Delta},
\end{equation}
where
\[
Q = 1 - \frac{x^2}{R_x^2 + \sigma} - \frac{y^2}{R_y^2 + \sigma} - \frac{z^2}{R_z^2 + \sigma},
\]
and
\begin{equation}
\Delta = \sqrt{(R_x^2 + \sigma)(R_y^2 + \sigma)(R_z^2 + \sigma)}.
\label{eq:Delta}
\end{equation}
Homoeoidal shells are shells situated inside the ellipsoid, bounded by equidensity surfaces of the density (\ref{rhos}). Using infinitesimally thin homoeoidal shells, the triple integral (\ref{phirho}) can be reduced to a single integral over the variable $\sigma$. For a detailed account on this integration process, see for instance references \cite{Routh, Chandrasekhar, Eberlein05}, or of course the original work by Ferrers \cite{Ferrers1877}. Notably, the right hand side of Eq. (\ref{phisn}) evaluates to a polynomial in $x, y,$ and $z$.

Next, Ferrers noted that whenever $\rho = 0$ at the boundary of the ellipsoid, then differentiation of the potential with respect to any of the cartesian coordinates, for example $x$, yields
\begin{equation}\label{dphidx}
\pdd{}{x}\phi[\rho](x,y,z) = \phi\left[\pdd{\rho}{x} \right](x,y,z),
\end{equation}
which can easily be checked with integration by parts. 
Finally, he noted that any monomial, such as $\rho_{ijk}$ defined in Eq.\ (\ref{rhoijk}), can be expressed by means of a series of differential coefficients of powers $s^m$ of the variable $s$ defined in (\ref{shomoeoid}), 
\begin{equation}\label{xiyjzk}
\rho_{ijk} = \sum_{m}\sum_{p+q+r\leq m} A^{(ijk)}_{mpqr} \frac{\mathrm{d}^{p + q + r}}{\mathrm{d}x^p \mathrm{d}y^q \mathrm{d}z^r} s^{m},
\end{equation}
where the $A^{(ijk)}_{mpqr}$ are constants and whereby it should be noted that the order of differentiation never exceeds $m$. By virtue of the latter observation, we can calculate the potential of the above density
by repeatedly applying the step (\ref{dphidx}) in the opposite direction, transferring all differential operators appearing in (\ref{xiyjzk}), from inside the integral $\phi[\rho_{ijk}]$ to the outside, since at each step we are ensured that the density being integrated over contains at least a factor of $s$, and hence is always equal to $0$ on the boundary. Thus, we arrive at the following result
\[
\phi[\rho_{ijk}](x,y,z) = \phi\left[\sum_{mpqr} A^{(ijk)}_{mpqr} \frac{\mathrm{d}^{p + q + r}}{\mathrm{d}x^p \mathrm{d}y^q \mathrm{d}z^r} s^{m}\right](x,y,z)
\]
\[ 
= \sum_{mpqr} A^{(ijk)}_{mpqr} \frac{\mathrm{d}^{p + q + r}}{\mathrm{d}x^p \mathrm{d}y^q \mathrm{d}z^r}\phi\left[s^m\right](x,y,z),
\]
in which the $\phi[s^m]$ terms are known through Eq.\ (\ref{phisn}). Recalling that the potentials $\phi[s^m]$ are in fact polynomial in $x,y,$ and $z$, and hence also any differential quotient, we can conclude that the potential $\phi[\rho_{ijk}]$ is also a polynomial. Crucial point in the above derivation is the observation expressed in equation (\ref{xiyjzk}), that any monomial can be written as a series of differential coefficients of a function of the homoeoidal shell index variable $s$, which is specific to ellipsoids only. In different geometries, some monomial densities might yield polynomial potentials, but in general this is not the case.

It remains to determine the precise coefficients of this polynomial, a task undertaken by F.W. Dyson who found a compact and elegant general expression \cite{Dyson1891}. Through the efforts of Ferrers and Dyson, the triple integral of (\ref{phirho}) which depended on the coordinate $\rvec$, is reduced to a finite number of single integrals appearing in the coefficients of a polynomial only. 

Although Dyson's formula in principle solves the problem, it is not particularly suited to numerical computation because it still contains differential operators. We therefore employ results from a more recent paper by Levin and Muratov \cite{Levin71}, which computes the polynomial coefficients of the potential $\phi[\rho_{ijk}]$ explicitly. Levin and Muratov make use of generalised depolarisation factors defined as
\begin{equation}\label{Mlmn}
M_{lmn} = (2l-1)!!(2m-1)!!(2n-1)!! \frac{\kappa_x \kappa_y \beta_{lmn}}{2 R_z^{2(l+m+n-1)}},
\end{equation}
where $m, l, n = 0, 1, 2 \ldots$, and $\beta_{lmn}$ is defined in Eq.\ (\ref{Betalmn}). Next, we write the exponents of $\rho_{ijk} = x^i y^j z^k$ as
\[
i = 2\lambda + \delta_\lambda, j = 2\mu + \delta_\mu, k = 2\nu + \delta_\nu,
\]
with $\lambda, \mu, \nu$ positive integers such that the $\delta_\mu, \delta_\nu, \delta_\lambda$ are either $0$ or $1$ for $i, j, k$ even or odd (respectively), and define
\[
\sigma = \lambda + \mu + \nu + 1.
\]
In the particular case of calculating the dipolar potential we are interested in the second derivative with respect to $z$, rather than the potential $\phi[\rho_{ijk}]$ itself. Using the results of Levin and Muratov then, this quantity is given by
\begin{widetext}
\begin{equation}\label{ddzphiijk}
\pdn{}{z}{2} \phi[\rho_{ijk}](x,y,z) = \frac{2R_x^i R_y^j R_z^k}{4^{\sigma}}i! j! k! \sum_{p=0}^{\sigma}\sum_{q=0}^{\sigma-p}\sum_{r=1}^{\sigma-p-q}\frac{S_{pqr} (2r + \delta_\nu)(2r + \delta_\nu - 1)x^{2p + \delta_\lambda}y^{2q + \delta_\mu}z^{2r + \delta_\nu - 2}}{(\sigma-p-q-r)!(2p \delta_\lambda + 1)(2q\delta_\mu + 1)(2r\delta_\nu + 1)}\Gamma_{pqr}^{(i,j,k)},
\end{equation}
where
\[
\Gamma_{pqr}^{(i,j,k)} = \sum_{l=0}^\lambda\sum_{m=0}^{\mu}\sum_{n=0}^{\nu}\frac{S_{lmn}R_x^{2l + \delta_\lambda}R_y^{2m + \delta_\mu}R_z^{2n + \delta_\nu}}{(\lambda - l)!(\mu - m)!(\nu - n)!(2l\delta_\lambda + 1)(2m\delta_\mu + 1)(2n\delta_\nu + 1)}M_{l + p + \delta_\lambda, m + q + \delta_ \mu, n + r + \delta_ \nu},
\]
\end{widetext}
and
\[
S_{lmn} = \frac{(-2)^{l+m+n}}{(2l)!(2m)!(2n)!}.
\]

\end{section}

\begin{section}{Calculating the dipolar potential inside an inverted paraboloidal ellipsoid}
\label{AppPotential2}
In this appendix we calculate the dipolar potential inside an ellipsoidally shaped BEC with a density profile of the form
\begin{equation}
n(\rvec)=n_{0}\left(1-\frac{x^2}{R_{x}^2}-\frac{y^2}{R_{y}^2}-\frac{z^2}{R_{z}^2}\right)=n_{0}s
\label{eq:ellipsoid}
\end{equation}
where $n_{0}$ is the density in the center of the BEC. This is a particular case of the more general density profile $\rho(\rvec)=s^{n}$ 
discussed in Appendix \ref{AppPotential} above, see Eq.\ (\ref{rhos}). Comparing with Eq.\ (\ref{DensParabolic}) in the main part of the text, we see that density profile (\ref{eq:ellipsoid}) corresponds exactly to the static solutions discussed in Section \ref{SecStaticSolutions}.
Note that the results in this appendix generalize those found in Appendix A of reference \cite{Eberlein05} for the cylindrically symmetric case to a triaxial ellipsoid.

In the case $n(\rvec)=n_{0} s$ considered in this appendix, it is straightforward to see that the integral (\ref{phisn}) for the fictitious electrostatic potential $\phi(x,y,z)$ inside the ellipsoid can be expressed as
\begin{eqnarray}
\phi & = & \frac{n_{0}R_{x} R_{y}R_{z}}{2} \bigg\{\frac{1}{4} +x^{2} \frac{\partial}{\partial(R_{x}^{2})}+y^{2} \frac{\partial}{\partial(R_{y}^{2})}+z^{2} \frac{\partial}{\partial(R_{z}^{2})}  
 \nonumber \\ && +2x^{2}y^{2} \frac{\partial^{2}}{\partial(R_{x}^{2})\partial(R_{y}^{2})}+2x^{2}z^{2} \frac{\partial^{2}}{\partial(R_{x}^{2})\partial(R_{z}^{2})}
  \nonumber \\ && 
 +2y^{2}z^{2} \frac{\partial^{2}}{\partial(R_{y}^{2})\partial(R_{z}^{2})} +\frac{1}{3}x^{4} \frac{\partial^{2}}{\partial(R_{x}^{2})^{2}} \nonumber \\ &&+\frac{1}{3}y^{4} \frac{\partial^{2}}{\partial(R_{y}^{2})^{2}} +\frac{1}{3}z^{4} \frac{\partial^{2}}{\partial(R_{z}^{2})^{2}}\bigg\} I_{A}(R_{x},R_{y},R_{z}) \label{eq:phi}
\end{eqnarray}
where
\begin{equation}
I_{A}= \int_{0}^{\infty} \frac{\romand \sigma}{\Delta},
\end{equation}
with $\Delta$ defined in Eq.\ (\ref{eq:Delta}).
For a prolate (cigar-shaped) condensate, defined as being when $R_{z}>R_{y}>R_{x}$, we find
 \begin{equation}
I_{A}=\frac{2}{\sqrt{R_{z}^{2}-R_{x}^{2}}}F \left( \arccos \left[ \frac{R_{x}}{R_{z}}  \right]  \left\vert \frac{R_{z}^{2}-R_{y}^{2}}{R_{z}^{2}-R_{x}^{2}} \right. \right)
\label{eq:Iprolate}
\end{equation}
where $F(\theta \vert m)$ is an elliptic integral of the first kind whose properties are well known \cite{AbramowitzStegun}. 
In the opposite case of an oblate (pancake-shaped) condensate, $R_{y}>R_{x}>R_{z}$, then
\begin{equation}
I_{A}=\frac{2}{\sqrt{R_{y}^{2}-R_{z}^{2}}}F \left( \arccos \left[ \frac{R_{z}}{R_{y}}  \right]  \left\vert \frac{R_{y}^{2}-R_{x}^{2}}{R_{y}^{2}-R_{z}^{2}} \right. \right).
\label{eq:Ioblate}
\end{equation}
The cylindrically symmetric case of $R_{x}=R_{y}$ is given in Appendix A of reference \cite{Eberlein05}. Thus, the problem of calculating the electrostatic potential $\phi(\rvec)$ reduces to one of finding derivatives of elliptic integrals, both with respect to the argument $\theta$ and the parameter $m$. 
To evaluate the derivatives of $I_{A}$ needed in Eq.\ (\ref{eq:phi}) we shall make use of the results
\begin{eqnarray}
\frac{\partial}{\partial \theta} F(\theta\vert m) & = & \frac{1}{\sqrt{1-m \sin^{2}\theta}} \label{eq:dFdm} \\
\frac{\partial}{\partial m} F(\theta \vert m) & = & \frac{E(\theta\vert m)}{2m(1-m)}-\frac{F(\theta \vert m)}{2m} \nonumber \\  && - \frac{\sin 2 \theta}{4(1-m)\sqrt{1-m \sin^{2} \theta}} 
\end{eqnarray}
where $E(\theta \vert m)$ is an elliptic integral of the second kind \cite{AbramowitzStegun}, and 
\begin{eqnarray}
\frac{\partial}{\partial \theta} E(\theta\vert m) & = & \sqrt{1-m \sin^{2}\theta } \\
\frac{\partial}{\partial m} E(\theta \vert m) & = & \frac{E(\theta \vert m)-F(\theta \vert m)}{2m}. \label{eq:dEdp}
\end{eqnarray}

When the external polarizing field is aligned along the $z$-axis then the mean-field dipolar potential $\Phi_{\mathrm{dd}}(\rvec)$ is given by Eq.\ (\ref{PhiddElectrostatic})
\[
\Phi_{\mathrm{dd}}(\rvec)=-C_{\mathrm{dd}}\left(\frac{\partial^{2}}{\partial z^{2}} \phi(\rvec) +\frac{1}{3}n(\rvec)\right).
\]
Taking $\phi(\rvec)$ from Equation (\ref{eq:phi}) one has
\begin{eqnarray}
\frac{\partial^{2}}{\partial z^{2}}  \phi &  = & \frac{n_{0} R_{x} R_{y}R_{z}}{2} \bigg\{2 \frac{\partial}{\partial(R_{z}^{2})}+4 x^{2} \frac{\partial^{2}}{\partial (R_{x}^{2}) \partial (R_{z}^{2})} \nonumber \\ && +4 y^{2} \frac{\partial^{2}}{\partial (R_{y}^{2}) \partial (R_{z}^{2})} \nonumber \\ & & +4z^{2} \frac{\partial^{2}}{\partial (R_{z}^{2})^{2}} \bigg\}I_{A}(R_{x},R_{y},R_{z}).
\end{eqnarray}
Thus, the dipolar mean-field potential $\Phi_{\mathrm{dd}}(\rvec)$ inside the inverted parabola density profile (\ref{eq:ellipsoid}) is itself a quadratic function of position $(x,y,z)$, as given in Eq.\ (\ref{PhiddParabolic}),
\begin{eqnarray}
&&\Phidd(\rvec) = - g \edd n(\rvec) +\frac{3 g \edd n_0 \kappa_x \kappa_y}{2} \nonumber\\ 
&&\times \left[\beta_{001}-\br{\beta_{101}x^2 + \beta_{011}y^2+3\beta_{002}z^2} R_z^{-2} \right] \nonumber
\end{eqnarray}
where $\kappa_{i}\equiv R_{i}/R_{z}$, and the coefficients  $\beta_{ijk}$ defined in Eq. (\ref{Betalmn}) can be seen to be
\begin{eqnarray}
\beta_{001} & = & - 2 R_{z}^{3} \frac{\partial}{\partial (R_{z}^{2})}  I_{A}(R_{x},R_{y},R_{z}) \label{eq:betacoefficientsappendix1} \\
\beta_{101} & = & 4 R_{z}^{5} \frac{\partial^{2}}{\partial (R_{x}^{2})(R_{z}^{2})}  I_{A}(R_{x},R_{y},R_{z}) \\
\beta_{011} & = & 4 R_{z}^{5} \frac{\partial^{2}}{\partial (R_{y}^{2})(R_{z}^{2})}  I_{A}(R_{x},R_{y},R_{z}) \\
\beta_{002} & = & \frac{4}{3} R_{z}^{5} \frac{\partial^{2}}{\partial (R_{z}^{2})^2}  I_{A}(R_{x},R_{y},R_{z}). 
\end{eqnarray}
Using (\ref{eq:dFdm}--\ref{eq:dEdp}) to perform the required derivatives, an explicit expression for the dipolar mean-field potential $\Phi_{\mathrm{dd}}$ can be given in terms of elliptic integrals. 
In the prolate case, when $R_{z}>R_{y}>R_{x}$, then
\begin{widetext}
\begin{eqnarray}
\beta_{001} & = & - \frac{2}{\sqrt{1-\kappa_{x}^{2}}(1-\kappa_{y}^{2})} \bigg\{ E \left( \arccos [\kappa_{x}]   \left\vert \frac{1-\kappa_{y}^{2}}{1-\kappa_{x}^{2}}  \right.   \right) -F \left( \arccos [\kappa_{x}]   \left\vert \frac{1-\kappa_{y}^{2}}{1-\kappa_{x}^{2}}  \right.   \right) \bigg\} \\
\beta_{101} & = & 2\bigg\{\frac{\kappa_{y}}{\kappa_{x}}\frac{1}{(1-\kappa_{x}^{2})(\kappa_{y}^{2}-\kappa_{x}^{2})}-\frac{1+\kappa_{x}^{2}-2\kappa_{y}^{2}}{(1-\kappa_{x}^{2})^{3/2}(\kappa_{y}^{2}-\kappa_{x}^{2})(1-\kappa_{y}^{2})} E \left( \arccos [\kappa_{x}]   \left\vert \frac{1-\kappa_{y}^{2}}{1-\kappa_{x}^{2}}  \right.   \right) \nonumber \\ & & -\frac{1}{(1-\kappa_{x}^{2})^{3/2}(1-\kappa_{y}^{2})}F \left( \arccos [\kappa_{x}]   \left\vert \frac{1-\kappa_{y}^{2}}{1-\kappa_{x}^{2}}  \right.   \right) \bigg\} \\
\beta_{011} & = & 2\bigg\{-\frac{\kappa_{x}}{\kappa_{y}}\frac{1}{(1-\kappa_{y}^{2})(\kappa_{y}^{2}-\kappa_{x}^{2})}+\frac{1-2\kappa_{x}^{2}+\kappa_{y}^{2}}{\sqrt{1-\kappa_{x}^{2}}(\kappa_{y}^{2}-\kappa_{x}^{2})(1-\kappa_{y}^{2})^{2}} E \left( \arccos [\kappa_{x}]   \left\vert \frac{1-\kappa_{y}^{2}}{1-\kappa_{x}^{2}}  \right.   \right) \nonumber \\ & & -\frac{2}{\sqrt{1-\kappa_{x}^{2}}(1-\kappa_{y}^{2})^{2}}F \left( \arccos [\kappa_{x}]   \left\vert \frac{1-\kappa_{y}^{2}}{1-\kappa_{x}^{2}}  \right.   \right) \bigg\} \\
\beta_{002} & = & \frac{2}{3}\bigg\{\frac{\kappa_{x}\kappa_{y}}{(1-\kappa_{y}^{2})(1-\kappa_{x}^{2})}-\frac{2(2-\kappa_{x}^{2}-\kappa_{y}^{2})}{(1-\kappa_{x}^{2})^{3/2}(1-\kappa_{y}^{2})^{2}} E \left( \arccos [\kappa_{x}]   \left\vert \frac{1-\kappa_{y}^{2}}{1-\kappa_{x}^{2}}  \right.   \right) \nonumber \\ & & +\frac{(3-2\kappa_{x}^{2}-\kappa_{y}^{2})}{(1-\kappa_{x}^{2})^{3/2}(1-\kappa_{y}^{2})^{2}}F \left( \arccos [\kappa_{x}]   \left\vert \frac{1-\kappa_{y}^{2}}{1-\kappa_{x}^{2}}  \right.   \right) \bigg\}. 
\end{eqnarray}

Whilst in the oblate case when $R_{y}>R_{x}>R_{z}$ we have

\begin{eqnarray}
\beta_{001} & = & -\frac{2}{(1-\kappa_{x}^2)(1-\kappa_{y}^{2}) \kappa_{y} } \bigg\{ \kappa_{x}(1-\kappa_{y}^2)+ \kappa_{y}\sqrt{\kappa_{y}^2-1} \ E \left( \arccos \left[\frac{1}{\kappa_{y}}\right]   \left\vert \frac{\kappa_{y}^{2}-\kappa_{x}^2}{\kappa_{y}^{2}-1}   \right. \right) \bigg\}  \\
\beta_{101} & = & 2\bigg\{\frac{1}{\kappa_{x}\kappa_{y}}\frac{1+\kappa_{x}^{2}}{(1-\kappa_{x}^{2})^{2}}+\frac{\kappa_{x}^{2}-2\kappa_{y}^{2}+1}{(\kappa_{y}^{2}-\kappa_{x}^{2})(\kappa_{x}^{2}-1)^{2}\sqrt{\kappa_{y}^{2}-1}} E \left( \arccos \left[\frac{1}{\kappa_{y}}\right]   \left\vert \frac{\kappa_{y}^{2}-\kappa_{x}^2}{\kappa_{y}^{2}-1}  \right.   \right) \nonumber \\
& & +\frac{1}{(\kappa_{y}^{2}-\kappa_{x}^{2})(\kappa_{x}^{2}-1)\sqrt{\kappa_{y}^{2}-1}}F \left( \arccos \left[\frac{1}{\kappa_{y}}\right]   \left\vert \frac{\kappa_{y}^{2}-\kappa_{x}^2}{\kappa_{y}^{2}-1}  \right.   \right) \bigg\} \label{eq:bxoblate} \\
\beta_{011} & = & 2\bigg\{\frac{\kappa_{x}}{\kappa_{y}}\frac{1}{(\kappa_{x}^{2}-1)(\kappa_{y}^{2}-1)}+\frac{2\kappa_{x}^{2}-\kappa_{y}^{2}-1}{(\kappa_{y}^{2}-\kappa_{x}^{2})(\kappa_{x}^{2}-1)(\kappa_{y}^{2}-1)^{3/2}} E \left( \arccos \left[\frac{1}{\kappa_{y}}\right]   \left\vert \frac{\kappa_{y}^{2}-\kappa_{x}^2}{\kappa_{y}^{2}-1}  \right.   \right) \nonumber \\
& & -\frac{1}{(\kappa_{y}^{2}-\kappa_{x}^{2})(\kappa_{y}^{2}-1)^{3/2}}F \left( \arccos \left[\frac{1}{\kappa_{y}}\right]   \left\vert \frac{\kappa_{y}^{2}-\kappa_{x}^2}{\kappa_{y}^{2}-1}  \right.   \right) \bigg\}   \label{eq:byoblate} \\
\beta_{002} & = & \frac{2}{3}\bigg\{\frac{\kappa_{x}}{\kappa_{y}}\frac{(\kappa_{x}^2 \kappa_{y}^{2} -2 \kappa_{x}^2-3\kappa_{y}^2 +4)}{(\kappa_{x}^{2}-1)^{2}(\kappa_{y}^{2}-1)}+\frac{2(\kappa_{x}^{2}+\kappa_{y}^{2}-2)}{(\kappa_{x}^{2}-1)^{2}(\kappa_{y}^{2}-1)^{3/2}} E \left( \arccos \left[\frac{1}{\kappa_{y}}\right]   \left\vert \frac{\kappa_{y}^{2}-\kappa_{x}^2}{\kappa_{y}^{2}-1}  \right.   \right) \nonumber \\
& & -\frac{1}{(\kappa_{x}^{2}-1)(\kappa_{y}^{2}-1)^{3/2}}F \left( \arccos \left[\frac{1}{\kappa_{y}}\right]   \left\vert \frac{\kappa_{y}^{2}-\kappa_{x}^2}{\kappa_{y}^{2}-1}  \right.   \right) \bigg\}.
 \label{eq:betacoefficientsappendix2}
\end{eqnarray}

\end{widetext}

\end{section}



\begin{thebibliography}{99}
\bibitem{NoteDipoleMode}
{For rotating systems the directions of oscillation will differ from the trap axes, but these systems are not considered in this work.}
\end{thebibliography}
\end{document}